\documentclass[a4paper,fleqn,usenatbib]{mnras}

\usepackage{txfonts}

\usepackage[T1]{fontenc}
\usepackage{ae,aecompl}

\usepackage{multicol}
\usepackage{color}
\usepackage{graphicx}
\usepackage{wasysym}

\def\aj{AJ}                   
\def\araa{ARA\&A}             
\def\apj{ApJ}                 
\def\apjl{ApJ}                
\def\apjs{ApJS}               
\def\apss{Ap\&SS}             
\def\aap{A\&A}                
\def\aapr{A\&A~Rev.}          
\def\mnras{MNRAS}             
\def\pasp{PASP}               

\newcommand{\kms}{\,$\rm km\;s^{-1}$}
\newcommand{\magasec}{\,mag\,arcsec$^{-2}$}
\newcommand{\msunpcsq}{\,M$_{\odot}$\,pc$^{-2}$} 
\newcommand{\wifes}{\textsc{WiFeS}}
\newcommand{\wise}{\textit{WISE}}
\newcommand{\galex}{\textit{GALEX}}
\newcommand{\mass}{\textsc{2Mass}}

\newcommand{\hopcat}{\textsc{Hopcat}}
\newcommand{\hipass}{\textsc{Hipass}}

\newcommand{\alfalfa}{\textsc{Alfalfa}}
\newcommand{\highmass}{H\textsc{i}ghMass}
\newcommand{\bluedisk}{\textsc{Bluedisk}}
\newcommand{\hix}{\textsc{Hix}}
\newcommand{\tirific}{\textsc{TiRiFiC}}

\newcommand{\pywifes}{\textsc{PyWiFeS}}
\newcommand{\hi}{H\,\textsc{i}}
\newcommand{\htwo}{H$_2$}
\newcommand{\kband}{\textit{K}-band}

\newcommand{\iband}{\textit{I}-band}
\newcommand{\rband}{\textit{R}-band}
\newcommand{\bjband}{$B_j$-band}
\newcommand{\nuv}{\textit{NUV}} 
\newcommand{\fuv}{\textit{FUV}}

\newcommand{\wthreeband}{\textit{W3}-band}

\newcommand{\gal}{ESO075-G006}

\title[The HIX galaxy survey I]{The HIX galaxy survey I: Study of the most gas rich galaxies from HIPASS}
\author[Lutz et al.]{K. A. Lutz$^{1, 2}$\thanks{E-mail: klutz@swin.edu.au},  V. A. Kilborn$^{1}$, B. Catinella$^{3}$, B. S. Koribalski$^{2}$, T. H. Brown$^{1, 3}$, \newauthor 
L. Cortese$^{3}$, H. D\'enes$^{2, 4}$, G. I. G. J\'ozsa$^{5, 6, 7}$ and O. I. Wong$^{3, 8}$ \\
$^{1}$ Centre for Astrophysics and Supercomputing, Swinburne University of Technology, P.O. Box 218, Hawthorn, VIC 3122, Australia \\
$^{2}$ Australia Telescope National Facility, CSIRO, P.O. Box 76, Epping, NSW 1710, Australia\\
$^{3}$ International Centre for Radio Astronomy Research (ICRAR), M468, The University of Western Australia, 35 Stirling Highway, \\
Crawley, WA 6009, Australia \\
$^{4}$ Research School of Astronomy \& Astrophysics, The Australian National University, Canberra, ACT 2611, Australia \\
$^{5}$ SKA South Africa Radio Astronomy Research Group, 3rd Floor, The Park, Park Road, Pinelands 7405, South Africa \\ 
$^{6}$ Rhodes Centre for Radio Astronomy Techniques \& Technologies, Department of Physics and Electronics, Rhodes University, \\
PO Box 94, Grahamstown 6140, South Africa \\ 
$^{7}$ Argelander-Institut f\"ur Astronomie, Auf dem H\"ugel 71, D-53121 Bonn, Germany \\
$^{8}$ ARC Centre of Excellence for All-Sky Astrophysics (CAASTRO)}

\date{Accepted XXX. Received YYY; in original form ZZZ}

\pubyear{2016}

\begin{document}
\label{firstpage}
\pagerange{\pageref{firstpage}--\pageref{lastpage}}
\maketitle

\begin{abstract}
We present the \hi\ eXtreme (\hix) galaxy survey targeting some of the most \hi\ rich galaxies in the southern hemisphere. The 13 \hix\ galaxies have been selected to host the most massive \hi\ discs at a given stellar luminosity. We compare these galaxies to a control sample of average galaxies detected in the \hi\ Parkes All Sky Survey (\hipass, \citealp{Barnes2001}). As the control sample is matched in stellar luminosity, we find that the stellar properties of \hix\ galaxies are similar to the control sample. Furthermore, the specific star formation rate and optical morphology do not differ between \hix\ and control galaxies. We find, however, the \hix\ galaxies to be less efficient in forming stars. For the most \hi\ massive galaxy in our sample (\gal, log M$\rm _{HI}$  [M$_{\odot}$] = $(10.8 \pm 0.1)$) the kinematic properties are the reason for inefficient star formation and \hi\ excess. Examining the Australian Telescope Compact Array (ATCA) \hi\ imaging and Wide Field Spectrograph (\wifes) optical spectra of \gal\ reveals an undisturbed galaxy without evidence for recent major, violent accretion events. A tilted-ring fit to the \hi\ disc together with the gas--phase oxygen abundance distribution supports the scenario that gas has been constantly accreted onto \gal\ but the high specific angular momentum makes \gal\ very inefficient in forming stars. Thus a massive \hi\ disc has been built up. 
\end{abstract}


\section{Introduction}
Over the last two decades large, systematic surveys of the atomic hydrogen (\hi) content of galaxies have allowed us to study the relation between star formation, stellar and \hi\ content of galaxies. Scaling relations have shown that the \hi\ content of galaxies in the local universe is well correlated with the galaxies' stellar content \citep{Haynes1984,Catinella2010,Denes2014,Brown2015}. For this work, outliers towards the very \hi\ rich end of such scaling relations are examined aiming to investigate why these galaxies host a more massive \hi\ disc than average. Two scenarios are considered: these galaxies are either more efficient at accreting gas from their environment, or they are less efficient at converting available gas into stars than \hi\ poor galaxies (see also \citealp{Huang2014}). A combination of the two scenarios is also possible.  

The mechanisms through which galaxies can accrete gas and replenish their gas reservoir are broadly separated into two categories: gas rich mergers ('clumpy accretion') or accretion of gas from the halo or the intergalactic medium ('smooth accretion'). 

Gas-rich mergers may include both minor and major mergers. Major mergers have been found to be a less dominant channel of gas accretion both in observational work and simulations (e.\,g. \citealp{vandeVoort2011}, \citealp{Wang2013} and also \citealp{Sancisi2008, SanchezAlmeida2014} and references therein). A series of gas-rich minor mergers may contribute to the gas replenishment of galaxies but observations have shown that also this channel does not deliver enough gas to a galaxy to sustain star formation \citep{DiTeodoro2014}.

In the local universe indirect evidence of cosmological gas accretion has been found: the integrated metallicity of galaxies is lower than expected in chemical evolution models of a closed and isolated galaxy (\citealp{vandenBergh1962}, \citealp{SanchezAlmeida2014} and references therein). Hence, the gas content of a galaxy needs to be replenished with metal-poor gas, i.\,e. gas originating from the intergalactic medium. This is supported by cosmological hydrodynamical simulations by \citet{Dave2013}, who among others have shown that the cycle of gas inflow, gas outflow and star formation is essential to reproduce scaling relations between two or three properties of galaxies, such as the gas content, the stellar content, the star formation activity or the metallicity. In particular the stellar mass-metallicity relation \citep{Hughes2013,Mannucci2010} is driven by smooth, stochastic accretion: the accretion of cosmological gas triggers star formation (increase of star formation) and dilutes the interstellar medium (decrease of the gas phase metallicity). Furthermore, \citet{Moran2012} find that massive galaxies with a higher \hi\ content have a steeper drop in gas-phase oxygen abundance. Putting this in the context of cosmological gas accretion, it implies that the outskirts of a galaxy are diluted by metal-poor inflowing gas. It is, however, suggested that in the local universe cosmological accretion is less efficient than at higher redshifts, as the \textsc{Halogas} survey was not able to detect the necessary amount of \hi\ clouds to sustain star formation \citep{Heald2011,Heald2015}.

In addition to cosmological gas accretion the 'Galactic Fountain' mechanism can facilitate the accretion of hot halo gas onto the galaxy \citep{Fraternali2011, Oosterloo2007}. In this model, the supernova driven galactic fountain initially expels metal-rich gas from the galactic disc into the halo. This now extra-planar gas cools down and falls back onto the disc. As this happens hot halo gas condenses onto the expelled gas, cools and gets dragged along into the disc. Thus, the gas content of the disc increases \citep{Marinacci2010,Marinacci2011}. In \hi\ observations the expelled, extra-planar gas can be identified by its lag in rotation velocity. An example for this process could be NGC\,891 \citep{Fraternali2011, Oosterloo2007}, but also several galaxies observed in the context of the \textsc{Halogas} survey \citep{Heald2011,Gentile2013} have shown evidence for lagging extra-planar gas. It is to be noted though that evidence for sufficient density of hot haloes is scarce (see e.\,g. \citealp{Bogdan2015}).

The alternative is that these very gas rich galaxies are inefficient at converting their available gas into stars. \citet{Wong2016} suggest that the \hi\ based star formation efficiency (SFE) is in particular related to the stability of a disc. An elevated specific angular momentum can be one mechanism to stabilise a disc towards gravitational instabilities \citep{Krumholz2016,Forbes2014i} and we therefore consider it as a reason for decreased SFE. An enhanced angular momentum has two effects on the galaxy. One is keeping recently accreted gas at larger galactocentric radii and lower densities \citep{Kim2013,Forbes2014i}, where conditions are less favourable for star formation. Secondly, a high angular momentum stabilises the gas against Jeans instabilities and subsequent star formation \citep{Toomre1964,Obreschkow2016}. Simulations suggest that the angular momentum in a local galaxy is independent of the angular momentum of the host halo \citep{Stevens2016}, but is determined by the merger history  and feedback \citep{Genel2015,Lagos2016}. Furthermore, the angular momentum is thought to constrain the upper envelope of the gas fraction - stellar mass plane \citep{Maddox2015}. 

In this paper we present the \hi\ eXtreme (HIX) galaxy survey. For this survey we are targeting galaxies that contain at least 2.5 times more \hi\ than expected from their stellar luminosity. In order for these galaxies to accumulate that much \hi\ we expect them to have either gone through a recent episode of elevated gas accretion (both smooth and clumpy), to be unable to form stars efficiently from their gas reservoir or a combination of these two scenarios.   

The \hix\ and a control sample of galaxies are observed in \hi\ with the Australian Telescope Compact Array (ATCA) and optical integral field spectra are obtained with the \wifes\ spectrograph for the \hix\ galaxies. This combined data set plus publicly available data allow to search for evidence of gas accretion as well as inefficient star formation as a reason for the \hi\ excess. 

The term \hi\ excess has been used in many different ways. We use the term ``\hi\ excess'' to describe a more massive \hi\ content than expected from optical properties. 

This article is structured as follow. In section \ref{sec:survey-sample}, we outline the sample selection and the survey strategy. In section \ref{sec:obs-data}, we present observed and publicly available data as well as the data reduction. Section \ref{sec:compare} compares our sample to the broader galaxy population. Section \ref{sec:focus} focuses on the most extreme galaxy in our sample, ESO075-G006, and in sections \ref{sec:discuss} and \ref{sec:conclude} we discuss our findings and conclude. 

Throughout the paper we will assume a flat $\Lambda$CDM cosmology with the following cosmological parameters: H$_0$ = 70.0\,km\,Mpc$^{-1}$\,s$^{-1}$, $\Omega_{m}$ = 0.3. All velocities are used in the optical convention (cz).

\section{Survey description and sample selection}
\label{sec:survey-sample}

\subsection{Sample selection}
The sample of \hix\ galaxies is selected from a parent sample of 1796\,galaxies in the Southern hemisphere. This is a subset of the \hipass\ catalogue and the \hipass\ bright galaxy catalogue \citep{Meyer2004,Koribalski2004} including only \hi\ detections with reliable, single optical counterparts. This sample has been compiled by \citet{Denes2014} to obtain scaling relations between the stellar and the \hi\ content of galaxies. Using their scaling relation between the \rband\ luminosity as published in \hopcat\ \citep{Doyle2005} and the \hipass\ \hi\ mass, an \hi\ mass is predicted for each galaxy and compared to the \hi\ mass as measured from the integrated 21\,cm emission line in \hipass. 

\begin{figure}
\includegraphics[width=3.15in]{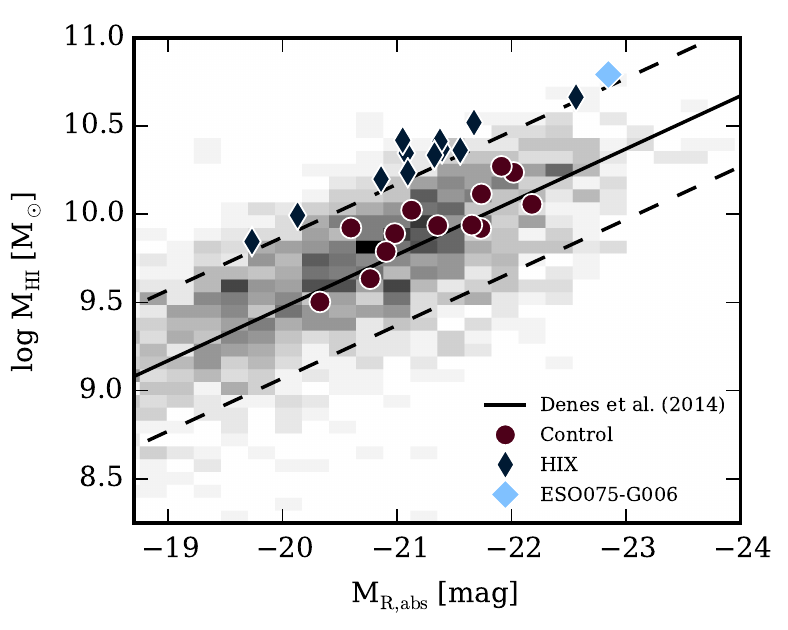}
\caption{The sample selection: The blue diamonds represent the \hix\ sample, the red circles the control sample. The light blue, wide diamond marks the data of galaxy \gal. The grey shade in the background represents the distribution of the \hipass\ parent sample. The black solid line gives the \citet{Denes2014} scaling relation and the black dashed line marks, where the measured \hi\ mass is 2.5\,times larger or smaller than the expected \hi\ mass.}
\label{fig:mhi_vs_magr}
\end{figure}

The detailed selection criteria for the \hix\ sample are: 
\begin{enumerate}
    \item an \hipass\ measurement of at least 2.5 times more \hi\ than expected from the scaling relation. Figure \ref{fig:mhi_vs_magr} shows the scaling relation by \citet{Denes2014} between the \rband\ absolute magnitude and the \hipass\  \hi\ mass, which was used here.
    \item absolute \kband\ magnitude $\rm M_{K} < -22.0$\,mag, restricting the sample to massive spiral galaxies. This is equivalent to a \kband\ luminosity $\rm log\,L_{K}  [L_{\odot}] = 42.8$ or a stellar mass of $\rm log\,M_{\star}  [M_{\odot}]= 9.7$.
    \item declination $< -30$\,deg for observability with the Australian Telescope Compact Array (ATCA). 
\end{enumerate}

We furthermore exclude galaxies near the galactic plane where the high density of foreground stars complicates the photometric measurements. 

SuperCOSMOS B-band images \citep{Hambly2001,Hambly2001i} and NED\footnote{http://ned.ipac.caltech.edu/} have been searched for neighbouring galaxies within the size of the \hipass\ spatial resolution element of 15.5\,arcmin \citep{Barnes2001}. As no galaxies of similar angular size and brightness have been found, \hi-excess due to source confusion can be excluded. Some galaxies, however, are accompanied by small dwarf galaxies. Basic properties of the \hi-excess galaxies are given in Table \ref{tab:hix_prop}. 

The parent sample for the \hix\ galaxies is a subset of \hipass. It is therefore already an \hi\ selected sample which is biased towards \hi\ rich systems. Selecting the outliers from this parent sample leads to the \hix\ galaxies being the most extreme galaxies with regards to their (relative) \hi\ mass in the Southern hemisphere.  

\begin{table*}
\begin{tabular}[H]{l || c c c c c c c c }
ID       & RA & DEC  & $\rm \log M_{\star} [M_{\odot}$] & $ \rm \log M_{HI} [M_{\odot}$] &  $\rm \log M_{HI} / M_{\star}$ & $\rm V_{sys} \ [km \ s^{-1}]$ & D [Mpc] & log SFR [$\rm M_{\odot}\,yr^{-1}$] \\
		(1)& (2) & (3) & (4) & (5) & (6) & (7) & (8) & (9) \\  \hline \hline
	HIX & ~& ~ &~ &~& ~& ~& ~ &\\ \hline
     ESO111-G014    & 00:08:18.8 & -59:30:56 & $ 10.5 $ & $ 10.6$ & $ 0.093$ & $7786$  & $112$ &  0.33 \\
     ESO243-G002    & 00:49:34.5 & -46:52:28 & $ 10.5 $ & $ 10.5$ & $-0.014$ & $8885$  & $129$ &  0.16 \\
       NGC\,0289    & 00:52:42.4 & -31:12:21 & $ 10.4 $ & $ 10.3$ & $-0.049$ & $1629$  & $23 $ & 0.081 \\
     ESO245-G010    & 01:56:44.5 & -43:58:21 & $ 10.4 $ & $ 10.3$ & $-0.075$ & $5740$  & $82 $ &  0.27 \\
     ESO417-G018    & 03:07:13.2 & -31:24:03 & $ 10.2 $ & $ 10.4$ & $ 0.14 $ & $4746$  & $67 $ &  0.18 \\
     ESO055-G013    & 04:11:43.0 & -70:13:59 & $ 10.1 $ & $ 10.4$ & $ 0.25 $ & $7386$  & $105$ &  0.21 \\
     ESO208-G026    & 07:35:21.1 & -50:02:35 & $ 9.79 $ & $ 9.89$ & $ 0.090$ & $2983$  & $40 $ & -0.50 \\
     ESO378-G003    & 11:28:04.0 & -36:32:34 & $ 9.96 $ & $ 10.2$ & $ 0.20 $ & $3021$  & $41 $ & -0.24 \\
     ESO381-G005    & 12:40:32.7 & -36:58:05 & $ 10.2 $ & $ 10.4$ & $ 0.21 $ & $5695$  & $80 $ &  0.22 \\
        IC\,4857    & 19:28:39.2 & -58:46:04 & $ 10.4 $ & $ 10.0$ & $-0.33 $ & $4679$  & $67 $ &  0.37 \\
     ESO461-G010    & 19:54:04.4 & -30:29:03 & $ 10.0 $ & $ 10.4$ & $ 0.40 $ & $6709$  & $98 $ & -0.13 \\
     ESO075-G006    & 21:23:29.5 & -69:41:05 & $ 10.5 $ & $ 10.8$ & $ 0.29 $ & $10609$ & $154$ &  0.46 \\
     ESO290-G035    & 23:01:32.5 & -46:38:47 & $ 10.4 $ & $ 10.3$ & $-0.11 $ & $5881$  & $84 $ & 0.012 \\   \hline
     CONTROL & ~& ~ &~ &~& ~& ~& ~ & \\ \hline
         NGC\,0685  & 01:47:42.8 & -52:45:42 &   $9.81$ & $9.53$  &  $-0.28$ & $1362$  & $18 $ &  -0.63\\
      ESO121-G026   & 06:21:38.8 & -59:44:24 &   $10.4$ & $9.95$  &  $-0.46$ & $2268$  & $30 $ &   0.12\\
      ESO123-G023   & 07:44:38.9 & -58:09:13 &   $10.1$ & $10.0$  &  $-0.06$ & $2914$  & $39 $ &  -0.54\\
      NGC\,3001     & 09:46:18.7 & -30:26:15 &   $10.8$ & $10.0$  &  $-0.81$ & $2464$  & $32 $ &   0.39\\
      ESO263-G015   & 10:12:19.9 & -47:17:42 &   $10.3$ & $9.67$  &  $-0.67$ & $2531$  & $33 $ &  -0.27\\
      NGC\,3261     & 10:29:01.5 & -44:39:24 &   $10.8$ & $10.1$  &  $-0.76$ & $2564$  & $34 $ &   0.26\\
      NGC\,4672     & 12:46:15.7 & -41:42:21 &   $10.7$ & $10.3$  &  $-0.43$ & $3273$  & $45 $ &   0.30\\
      NGC\,5161     & 13:29:13.9 & -33:10:26 &   $10.5$ & $10.1$  &  $-0.34$ & $2390$  & $32 $ &   0.14\\
      ESO383-G005   & 13:29:23.6 & -34:16:17 &   $10.4$ & $9.75$  &  $-0.67$ & $3614$  & $50 $ &  -0.18\\
      IC\,4366      & 14:05:11.5 & -33:45:36 &   $10.7$ & $10.1$  &  $-0.61$ & $4615$  & $65 $ &   0.54\\
      ESO462-G016   & 20:23:39.0 & -28:16:40 &   $10.1$ & $9.85$  &  $-0.23$ & $3055$  & $45 $ &  0.073\\
      ESO287-G013   & 21:23:13.9 & -45:46:23 &   $10.5$ & $10.2$  &  $-0.34$ & $2697$  & $39 $ &   0.18\\
      ESO240-G011   & 23:37:49.9 & -47:43:41 &   $10.9$ & $10.3$  &  $-0.55$ & $2840$  & $40 $ & -0.0013\\
\end{tabular}                                                                                           
\caption{The \hi\ extreme and control galaxy sample. \textit{Column} (1): An ID for every galaxy. \textit {Column} (2) and (3): RA and DEC taken from 2MASS. \textit{Column} (4): The stellar mass calculated as described in Section \ref{sec:basic_props}. \textit{Column} (5): The \hi\ mass calculated following Equation \ref{equ:himass} using the measured \hipass\ flux and the distance as stated in \textit{Column} (8). \textit{Column} (6): The \hi\ gas mass fraction is defined as $\rm f_{HI} = \frac{M_{HI}}{M_{\star}}$. \textit{Column} (7): The heliocentric, systemic velocity from \hipass\ \citep{Meyer2004}. Velocities here and throughout the paper are given in the optical convention (cz). \textit{Column} (8): Luminosity distance calculated from the systemic velocity in column 7 converted to the Galactic Standard of Rest frame and assuming our standard cosmology. \textit{Column} (9): The star formation rate measured as described in Section \ref{sec:basic_props}.}
\label{tab:hix_prop}                                                                                                                                                                                                                                                                                                                                                                                                                                                                                                                                                                                                                                                                                                                                                                                                                                                                                                                                                    
\end{table*}

\subsection{Control and comparison samples}
In order to compare the \hix\ sample to a sample of average \hipass\ galaxies, a control sample has been compiled. The control galaxies were drawn from the same parent sample and the selection criteria (ii) and (iii) were the same as used to define the \hix\ sample. However, criterion (i) was changed to: \\

(i) a measured \hipass\ \hi\ mass between 1.6 times less and more than what is expected from the scaling relation in order to capture galaxies well within the scatter (0.3\,dex). \\

The ATCA Online Archive\footnote{http://atoa.atnf.csiro.au/} (ATOA) was searched for observations of those galaxies that follow the control sample selection criteria. This lead to 13 control galaxies, that have been observed for at least one synthesis in any 1.5-km array configuration of ATCA. In Figure \ref{fig:mhi_vs_magr} these galaxies are drawn as red circles and lie well within the scatter of the scaling relation. More details and basic properties of the control sample are given in Table \ref{tab:hix_prop}. 

It is to be noted that the control sample is selected to be normal for an \hi\ selected sample as e.\,g. \hipass. This implies that the control sample is potentially still \hi\ rich compared to a stellar mass selected sample.

\section{Observations, data reduction and auxiliary data}
\label{sec:obs-data}

\subsection{ATCA observations}
In this work we only present the ATCA observations for \gal, which is the most extreme galaxy in our sample with respect to (relative) \hi\ mass and will be further discussed in Section \ref{sec:focus}. The ATCA \hi\ data of the remaining \hix\ and control galaxies will be presented in a subsequent paper. 

For the observations of \gal, the standard ATCA flux and bandpass calibrator PKS\,1934-638 has also been observed as the phase calibrator, which was visited regularly during the observations. The observations were carried out between November\,2012 and January\,2013. On source times are 5.6\,h, 9.3\,h and 5.6\,h in the 1.5D, 750C and EW367 array configurations respectively. The Compact Array Broadband Back end (CABB, \citealp{Wilson2011}) allows to observe continuum and spectral line simultaneously. The continuum band is 2048\,MHz wide, has a frequency resolution of 1\,MHz and is centred on 2.1\,GHz. Spectral line observations cover a 8.5\,MHz bandwidth with a channel width of 0.5\,kHz. This is equivalent to a velocity resolution of 0.1\,\kms\ at 1370\,MHz, which is the band's central frequency. In this paper we only utilise the spectral line data, which will be smoothed to channel widths of 4 and 10\kms.

\subsection{Radio data reduction and analysis}
The data reduction of the ATCA \hi\ observations is completed using standard tasks in the program package \textsc{miriad} \citep{Sault1995} and the python wrapper \textsc{mirpy}\footnote{https://pypi.python.org/pypi/mirpy}.

After removing radio frequency interference, a semi-automated pipeline conducts bandpass, flux and, phase calibration and subtracts a first order polynomial baseline for each array configuration separately. All available observations from different array configurations are then combined in the Fourier transformation. For the detailed analysis of the \hi\ properties of \gal\ in Section \ref{sec:focus}, two different data cubes were produced. The cube for the moment analysis has been combined by using a Brigg's robust parameter of 0.5 and a channel width of 4\kms\ to maximise sensitivity, the cube for the tilted ring fit with a robust parameter of 0.0 and a velocity channel width of 10\kms\ to maximise spatial resolution. As the longest baselines with Antenna\,6 are in most cases too long to pick up signal from diffuse and extended \hi\ and merely add noise to the data, all baselines with Antenna\,6 are excluded. For the \hi\ data of \gal, this leads to synthesised beam sizes of 52.4\,arcsec by 41.2\,arcsec (rob=0.5) and 40.6\,arcsec by 25.3\,arcsec (rob=0.0) and a noise level of 1.81\,mJy\,beam$^{-1}$ and 1.2\,mJy\,beam$^{-1}$ respectively. The lowest column density limit in the moment analysis (on rob=0.5 data) assumes a 3\,$\sigma$ detection over a velocity width of 12\kms and amounts to $\rm 70\,mJy\,beam^{-1}\,km\,s^{-1} = 0.4 \cdot 10^{20}\,cm^{-2} = 0.3\,M_{\odot}\,pc^{-2}$. Moment maps have been created with the \textsc{miriad} task \textit{moment}, including a 3\,$\sigma$ clipping. Moment 1 and 2 maps are further masked to the lowest column density contour and all moment maps are regridded to the resolution of the SuperCOSMOS optical image.

The resulting ATCA data cubes are then further analysed with standard \textsc{miriad} tasks, SAOImage DS9, \tirific\ \citep{Jozsa2007} and python scripts \footnote{https://www.python.org/}. 

\subsection{WiFeS observations}
\gal\ has also been observed with the Wide Field Spectrograph (\wifes, \citealp{Dopita2007}) integral field unit (IFU) mounted on the Siding Spring Observatory ANU 2.3\,m telescope. \gal\ has been observed with three pointings towards the north-western edge, the centre and the south-eastern edge of the disc. \wifes\ has been setup with $R=3000$ gratings in both the red and blue arm in combination with the dichronic set at 560\,nm. For reliable skyline subtraction, galaxy data have been taken in the nod-and-shuffle mode. Every night of observations has been completed with standard calibration images including bias, sky and dome flat field and wire imaging for centring the slitlets as well as NeAr and CuAr arc lamp spectra for wavelength calibration. Two to three times a night a standard star for flux calibration has been observed. The observations of \gal\ have been conducted in August 2014, July and October 2015 under photometric conditions. For more details of the three pointings see Table \ref{tab:wifes}. 

\begin{table}
\begin{tabular}[H]{l c c c c}
Pointing & RA & DEC & PA & $\rm t_{exp}$ \\ 
(1) & (2) & (3) & (4) & (5) \\ \hline
Centre & 21:23:28.7 & -69:41:02 & $165\,\degr$ & 1800s \\
South  & 21:23:30.7 & -69:41:31  & $ 75\,\degr$ & 3600s \\
North  & 21:23:26.9 & -69:40:32  & $ 75\,\degr$ & 1800s \\

\end{tabular}                                                                                           
\caption{Details for the three \wifes\ pointings on \gal. \textit{Column} (1): an ID for the pointing. \textit{Column} (2) and (3): the central coordinates. \textit{Column} (4): the rotator angle of the spectrograph. \textit{Column} (5): the total on-source exposure time. }
\label{tab:wifes}                                                                                                                                                                                                                                                                                                                                                                                                                                                                                                                                                                                                                                                                                                                                                                                                                                                                                                                                                    
\end{table}

\subsection{WiFeS data reduction and analysis}
\citet{Childress2014} provide the fully automated \pywifes\ pipeline for \wifes\ data. This pipeline includes bad pixel repair, bias  and dark current subtraction, flat fielding, wavelength calibration, sky subtraction, flux calibration and data cube creation. The resulting data cubes are 70 by 25 pixels (35 by 25\,arcsec) in size.

\pywifes\ is run on each observed galaxy data cube individually. For each pointing, data cubes are median stacked afterwards. This way small spatial offsets between observations can be taken into account. This procedure results in two data cubes for each pointing: one containing the red half of the spectrum and the other one the blue half. 

One stellar population model is constructed for each pointing by median stacking the spectra of all spaxels in one pointing. The stellar population fit is conducted on both the red and blue cube together using the pPXF method and complimentary Python script by \citet{Cappellari2004}. The stellar population model library is taken from \citet{Vazdekis2010}.

For the measurement of the emission line strength the data cubes of the red and blue half of the spectrum are analysed independently. Each individual data cube is Voronoi binned following \citet{Cappellari2003,Cappellari2009}. The Voronoi binning is constructed such that the signal to noise ratio in the wavelength range around the O[III] (blue) or N[II] line (red) reaches 50. The error map necessary for the binning is constructed from the variance data cube as given by the \pywifes\ pipeline. The size of final Voronoi bins are between 1 and 200 spaxels. In each Voronoi bin the stellar continuum model as fitted to the entire pointing is subtracted. Emission lines are fitted with Gaussians in each Voronoi bin. This results in a redshift measurement from the H$\alpha$ line and line strengths that can be used to estimate the gas-phase oxygen abundance.

The Voronoi bins of the red and the blue arm are different. So each pixel is assigned the value of the entire bin and metallicities are calculated on a pixel by pixel basis. 

\subsection{Public Data}
In addition to the observed data, publicly available data are also used:
\begin{enumerate}
    \item The \hipass\ survey data are used to calculate the \hi\ mass \citep{Barnes2001,Meyer2004,Zwaan2004}. 
    \item The AllWISE data release \citep{Cutri2013} of the \textit{WISE} mission \citep{Wright2010} provides imaging in the four mid-infrared \textit{WISE} bands: W1 (3.4\,$\mu$m), W2 (4.6\,$\mu$m), W3 (12\,$\mu$m), W4 (22\,$\mu$m). The flux of galaxies in each band is measured from images obtained from the ICORE co-adder\footnote{http://irsa.ipac.caltech.edu/applications/ICORE/} \citep{Masci2009} and converted into magnitudes following the prescription in \citet{Cutri2013}. 
    \item The catalogue of extended sources in the Two Micron All Sky Survey (2MASX, \citealp{Skrutskie2006}) provides magnitudes in the \textit{J}, \textit{H} and \kband\ and precise coordinates of the centre of the stellar disc. 
    \item The \hipass\ optical counter part catalogue (\hopcat, \citealp{Doyle2005}) provides \textit{B$_j$}, \textit{R} and \iband\ magnitudes, optical position angles and optical semi-major to semi-minor axis ratios from SuperCOSMOS \citep{Hambly2001,Hambly2001i}.  \hopcat\ also states the Galactic extinction measures $E(B-V)$ at the location of each galaxy as estimated by \citet{Schlegel1998}. 
    \item Optical imaging used in this work is obtained from the SuperCOSMOS \citep{Hambly2001,Hambly2001i} web page\footnote{http://www-wfau.roe.ac.uk/sss/index.html}.
    \item The ESO-LV catalogue \citep{Lauberts1989} provides both isophotal diameters and diameters at 90 per cent \textit{B} light
    \item Near- and far ultraviolet (NUV, FUV) photometry has been measured from the Galaxy Evolution Explorer imaging (\textit{GALEX}, \citealp{Martin2005, Morrissey2007}). \nuv\ images are available for all except one \hix\ galaxy (ESO208-G026) and except four control galaxies (NGC\,685, NGC\,3001 and NGC\,3261, ESO123-G023). \nuv\ counts are measured from intensity maps and converted into magnitudes using \galex 's flux calculator \footnote{http://asd.gsfc.nasa.gov/archive/galex/FAQ/counts\_background.html}.
\end{enumerate}

The \wise\ and \galex\ photometry is remeasured in elliptical apertures, where the semi major axis is of the size of the ESO-LV 25\magasec\ isophotal radius and the axis ratio is taken from \hopcat. Stars are masked out. The local sky background is estimated from dedicated background apertures and subtracted either with \textsc{funtools}\footnote{https://github.com/ericmandel/funtools} (GALEX) or as prescribed in the supplement (WISE).

The \mass\ data for the \hipass\ parent sample are retrieved from the VizieR catalogue access tool, which also took care of reliable cross-matching. 

\begin{figure*}
\begin{tabular}[H]{c c c c}
\includegraphics[width=0.22\textwidth]{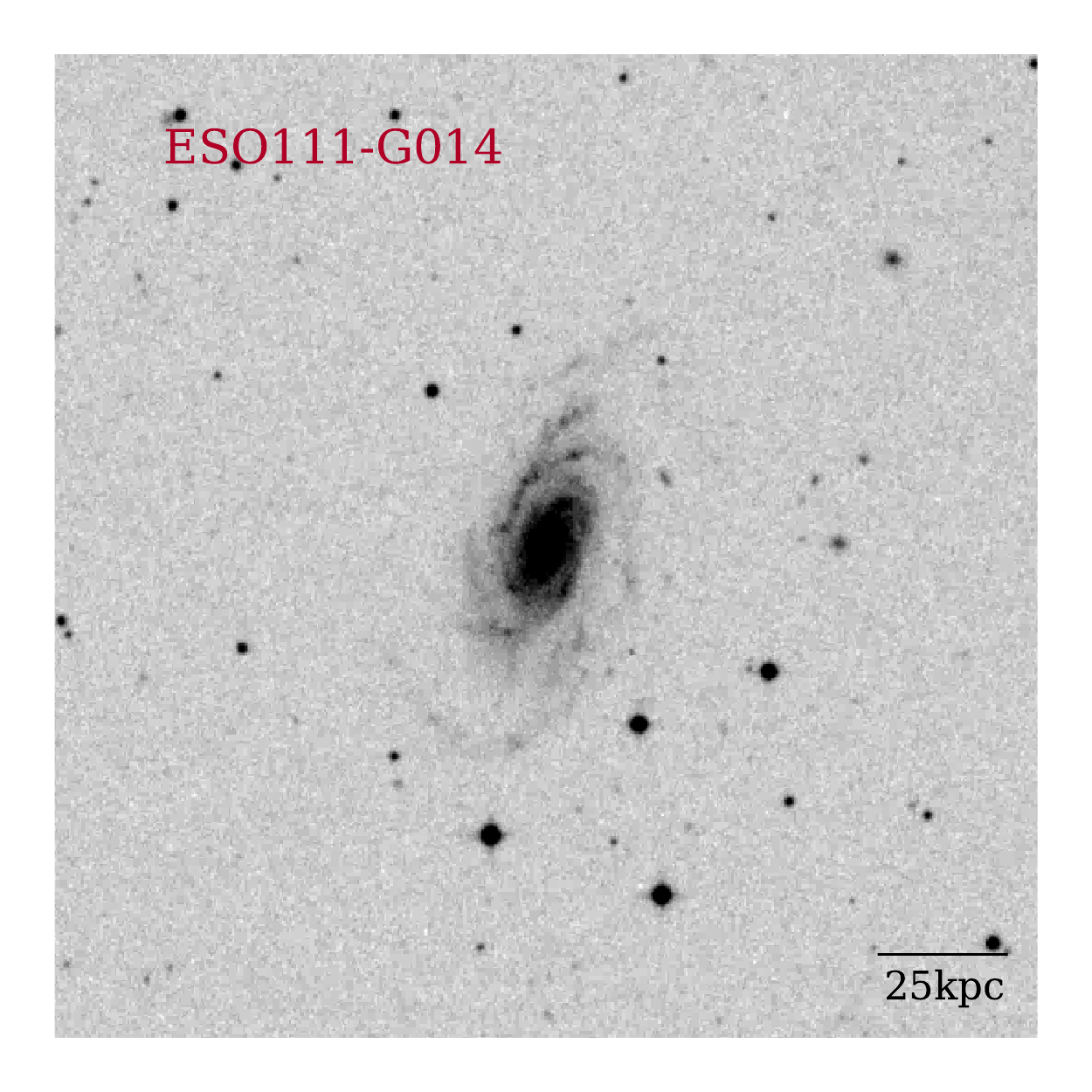} & \includegraphics[width=0.22\textwidth]{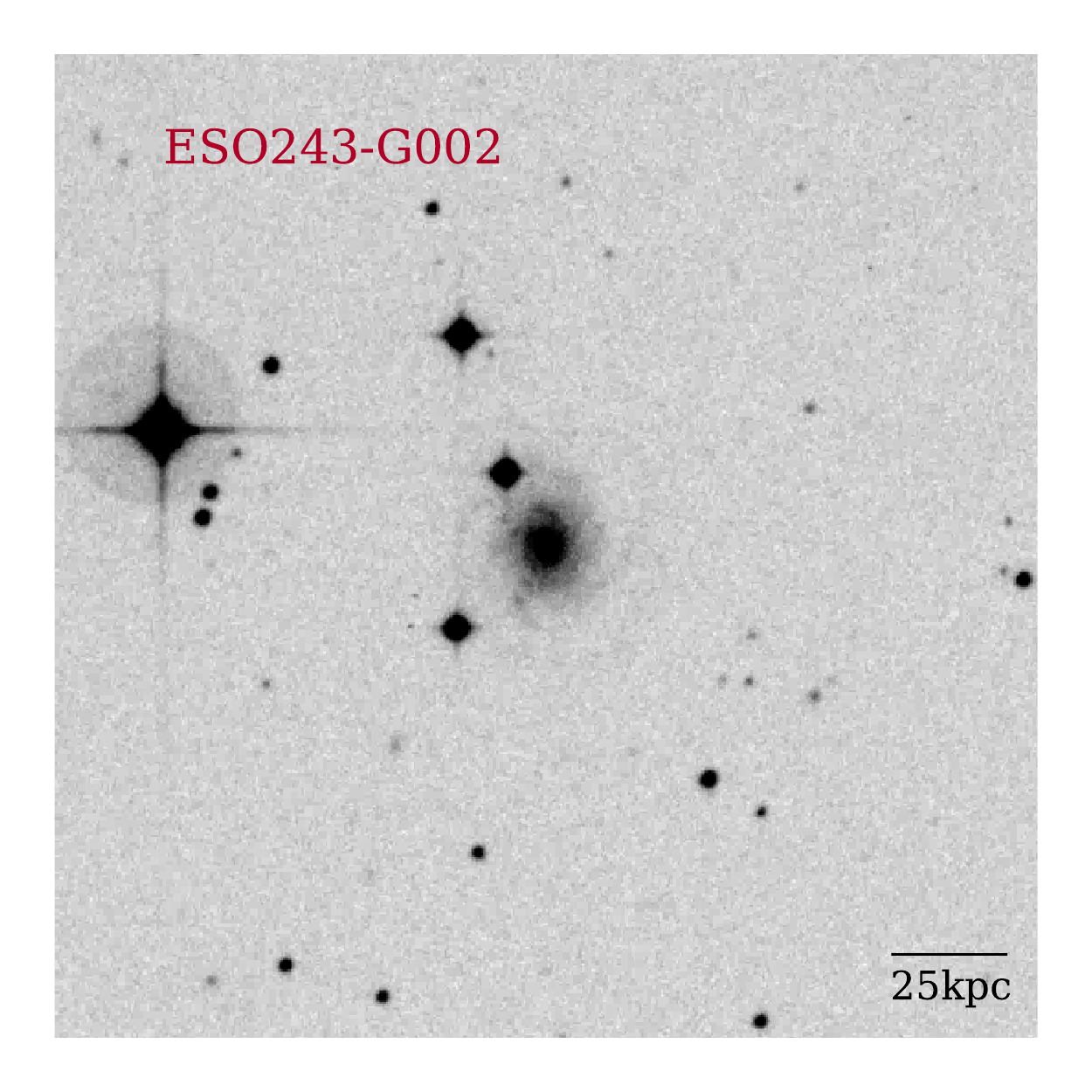} &
\includegraphics[width=0.22\textwidth]{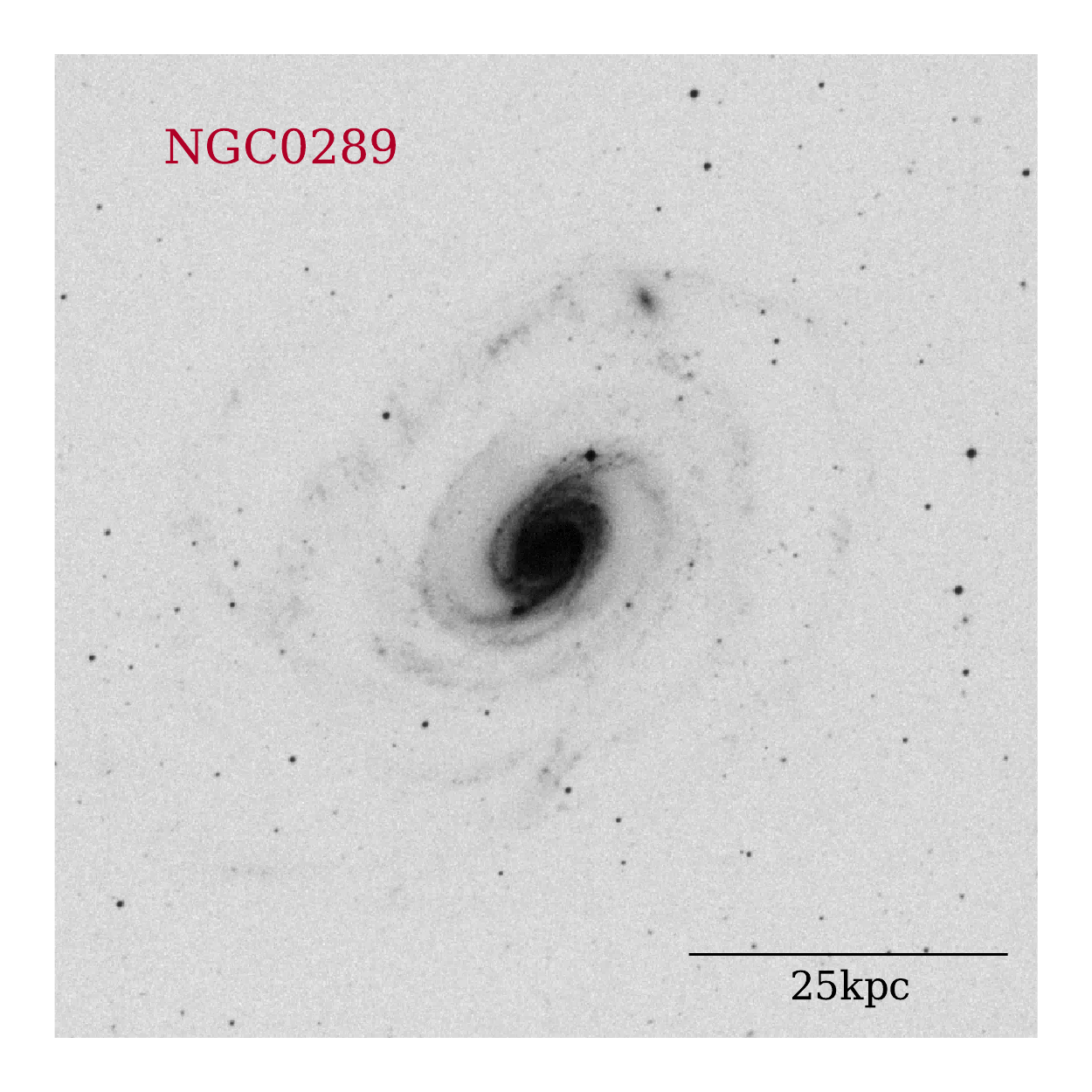} & \includegraphics[width=0.22\textwidth]{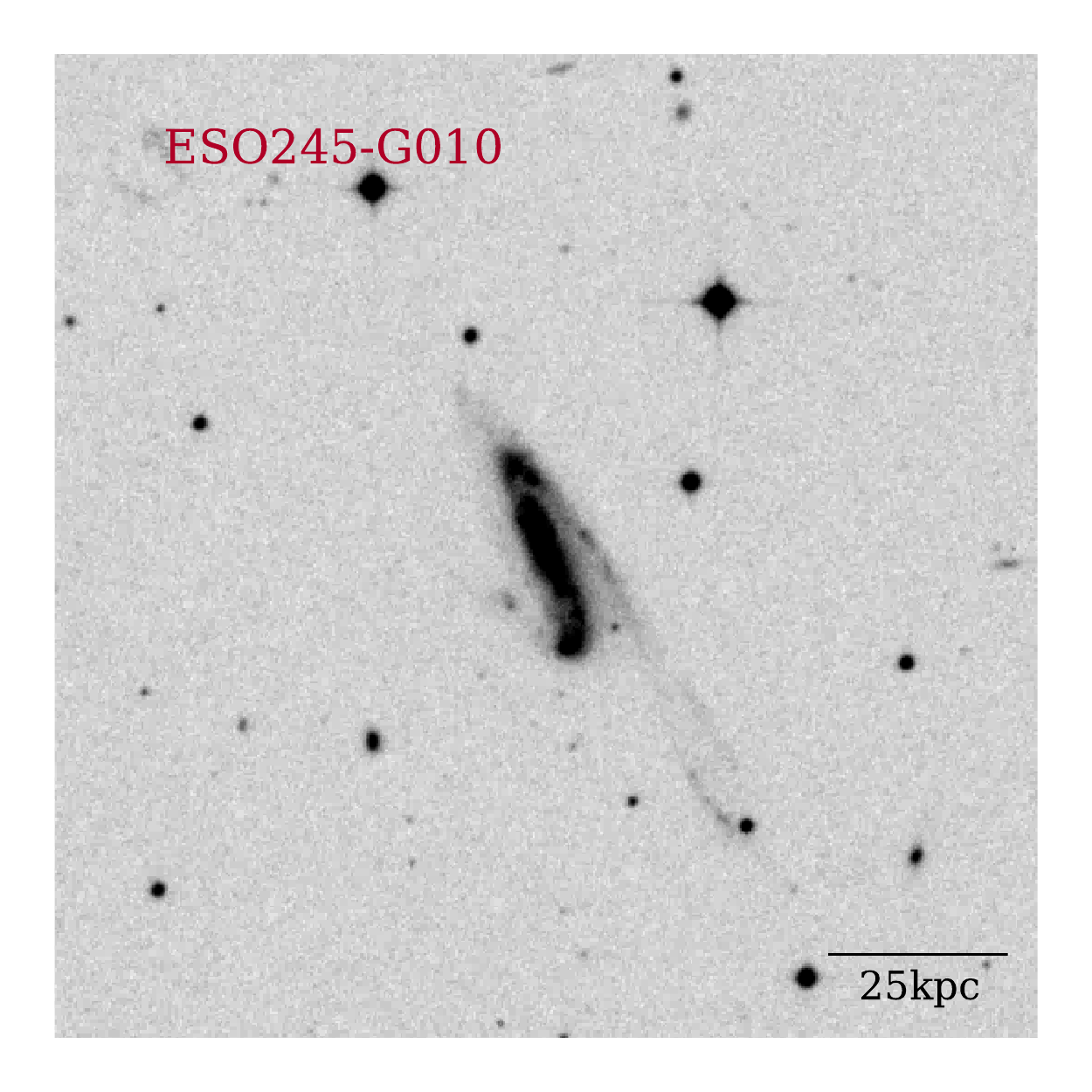} \\
\includegraphics[width=0.22\textwidth]{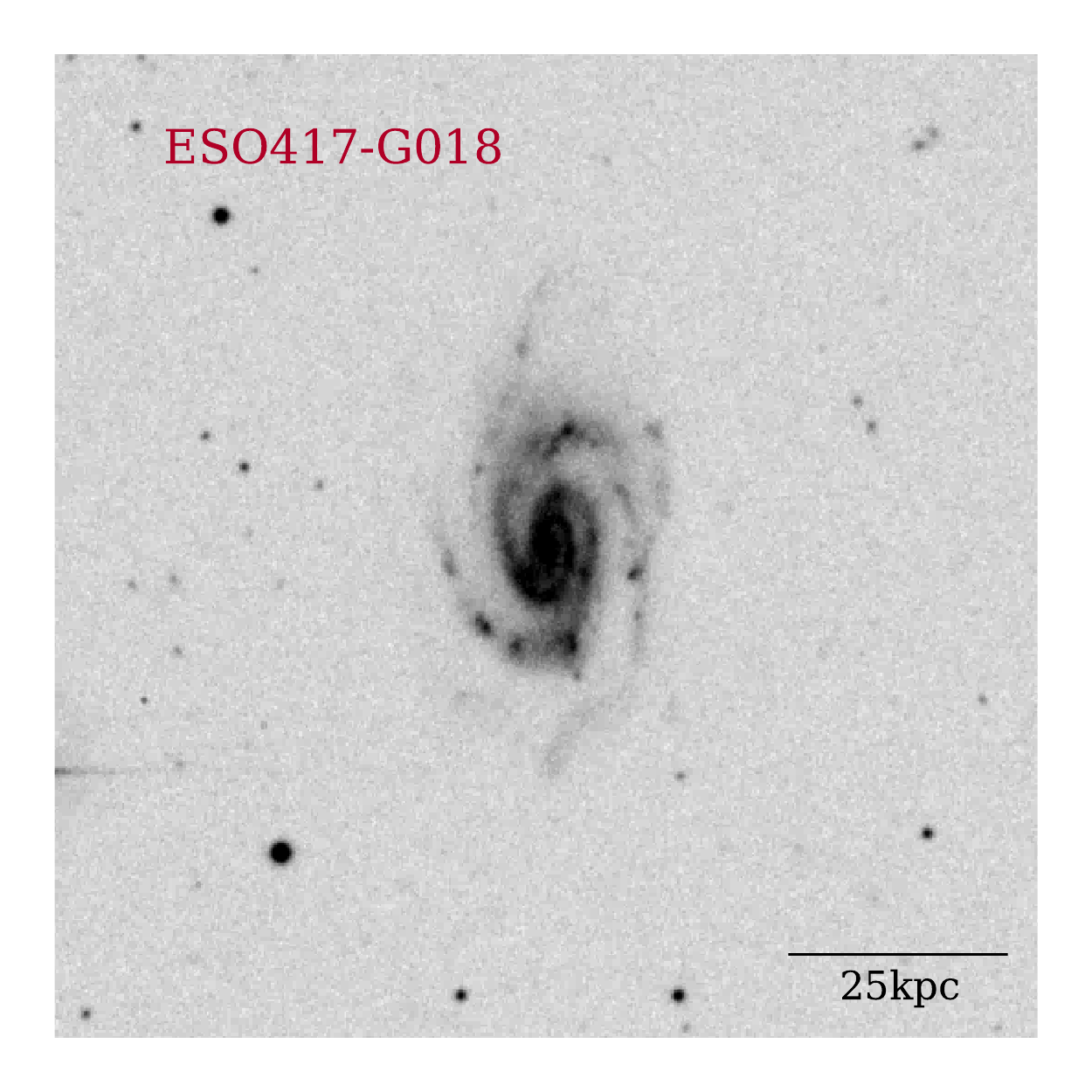} & \includegraphics[width=0.22\textwidth]{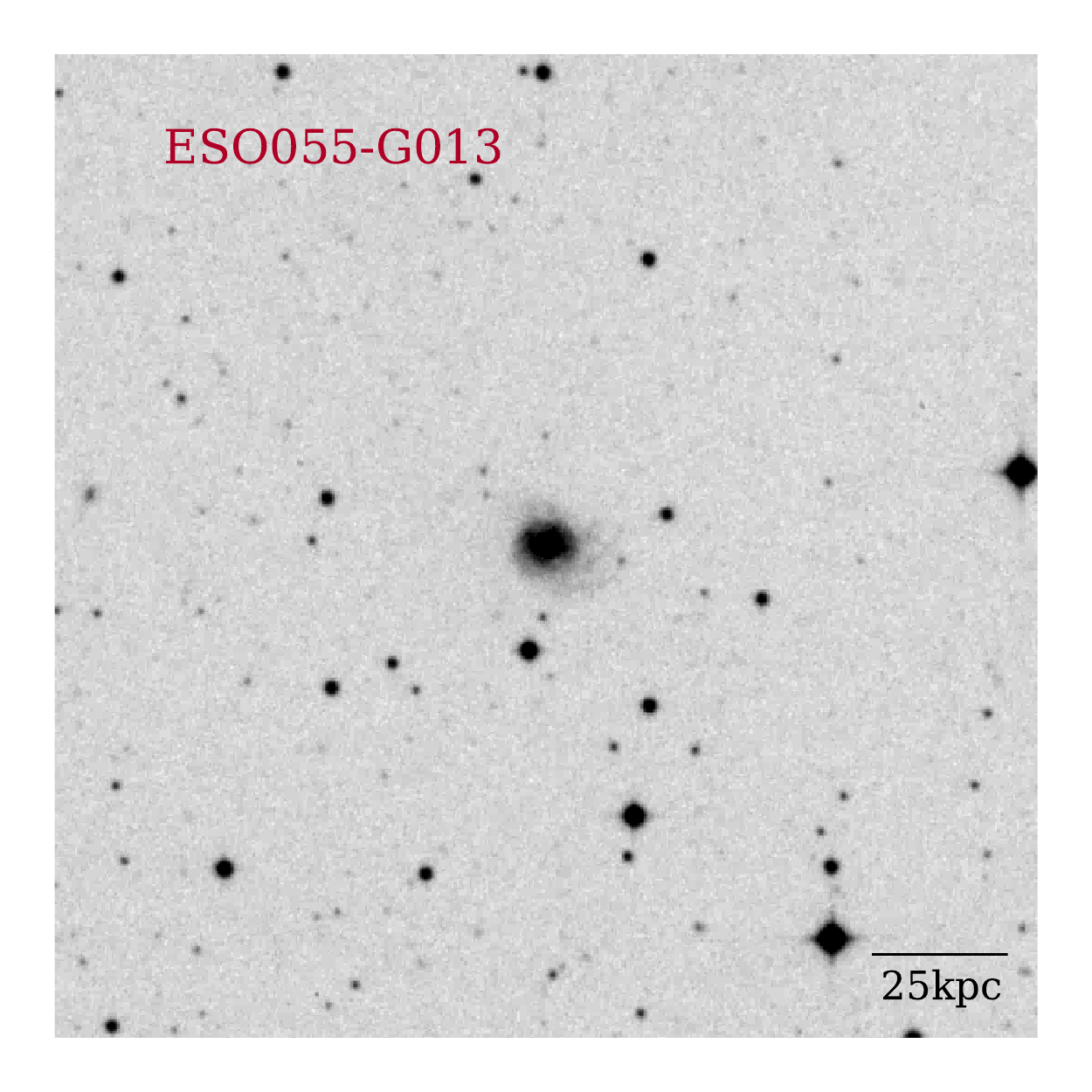} & 
\includegraphics[width=0.22\textwidth]{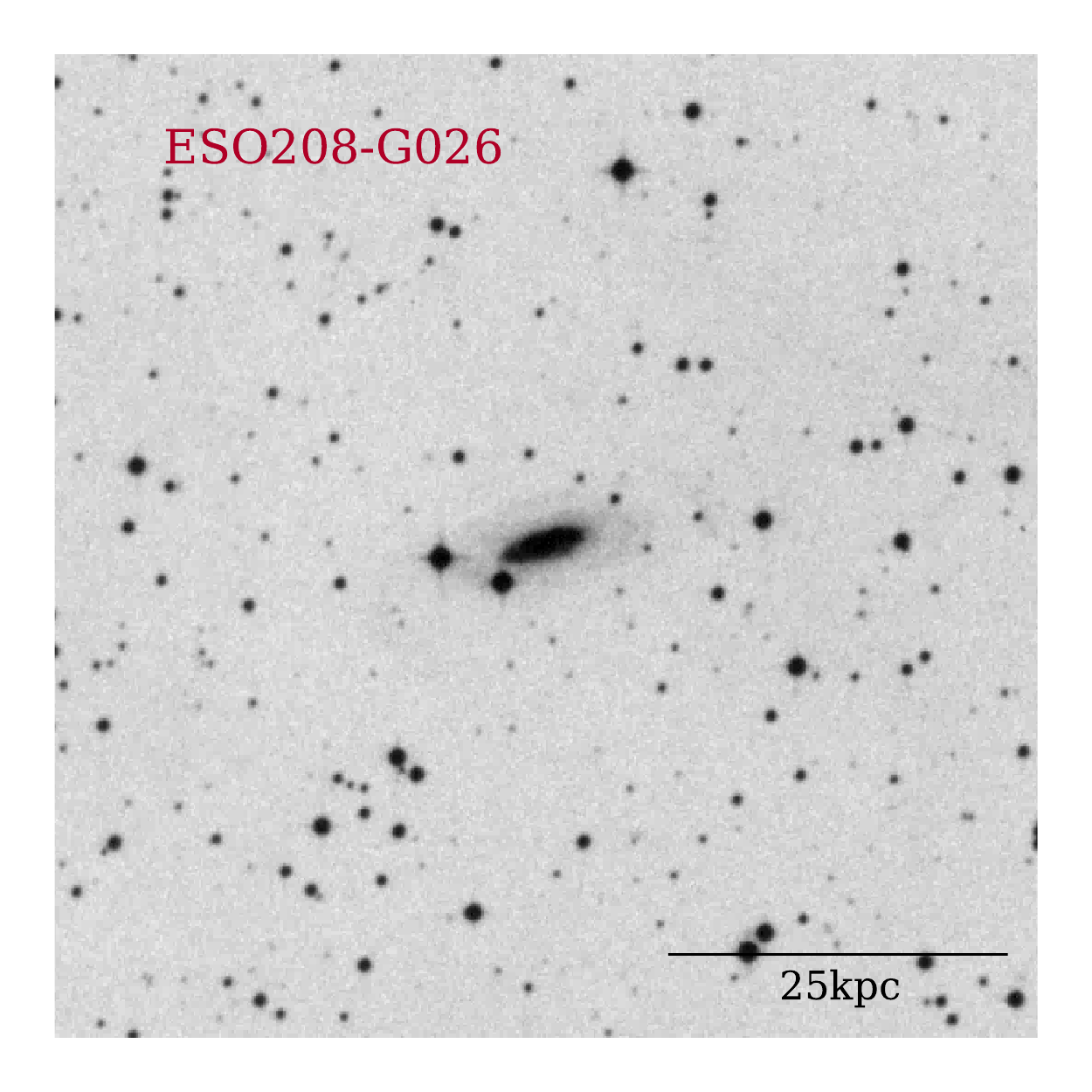} & \includegraphics[width=0.22\textwidth]{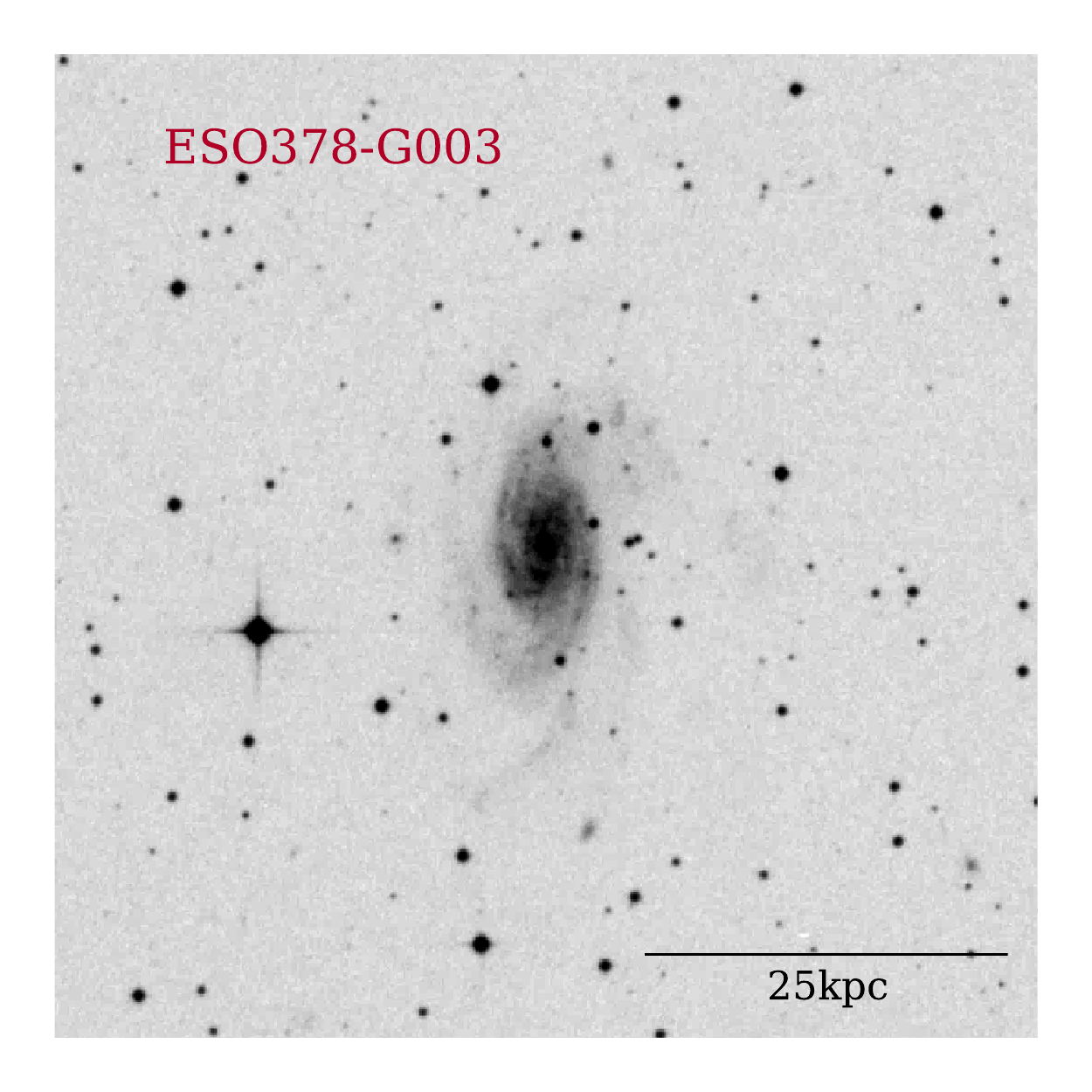} \\ 
\includegraphics[width=0.22\textwidth]{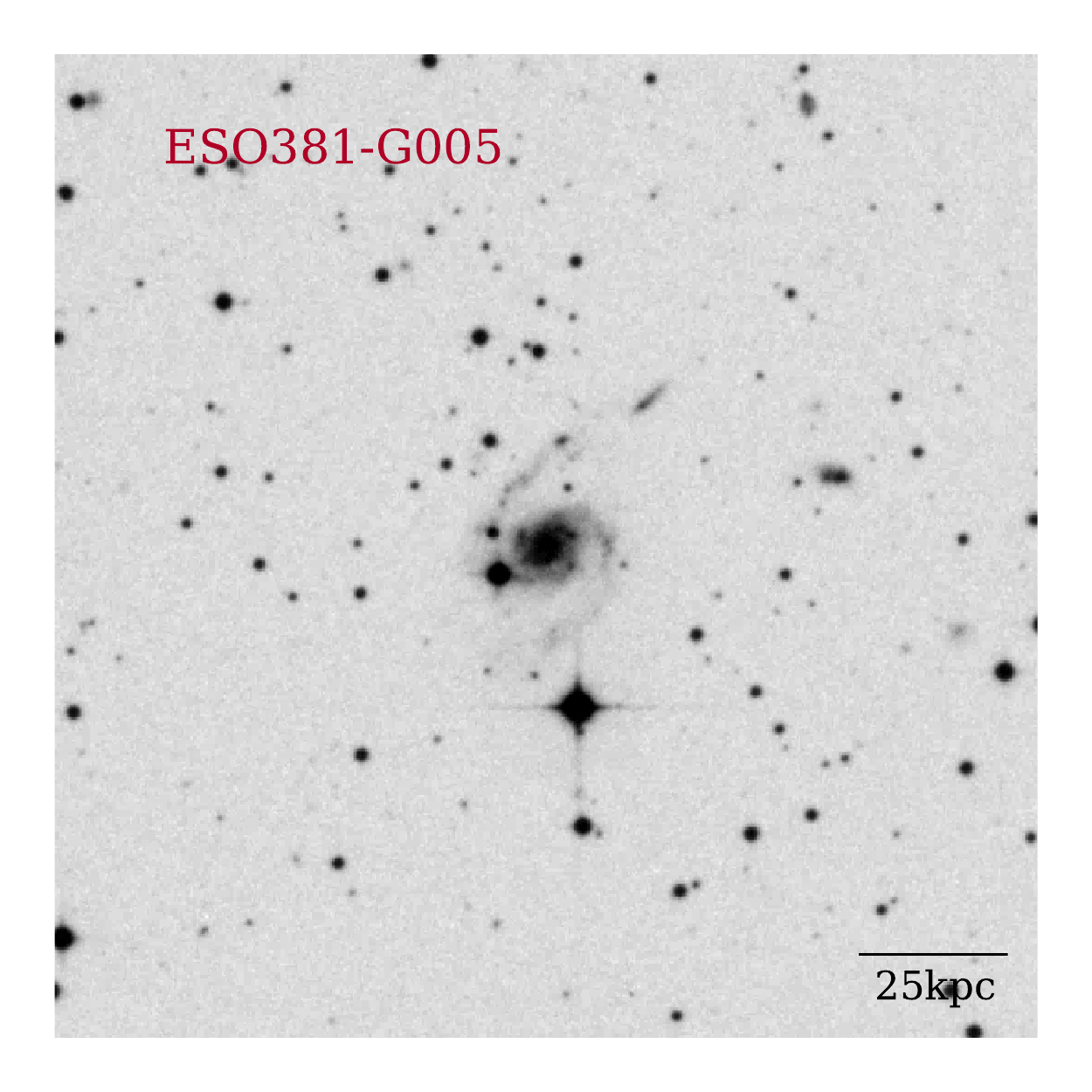} & \includegraphics[width=0.22\textwidth]{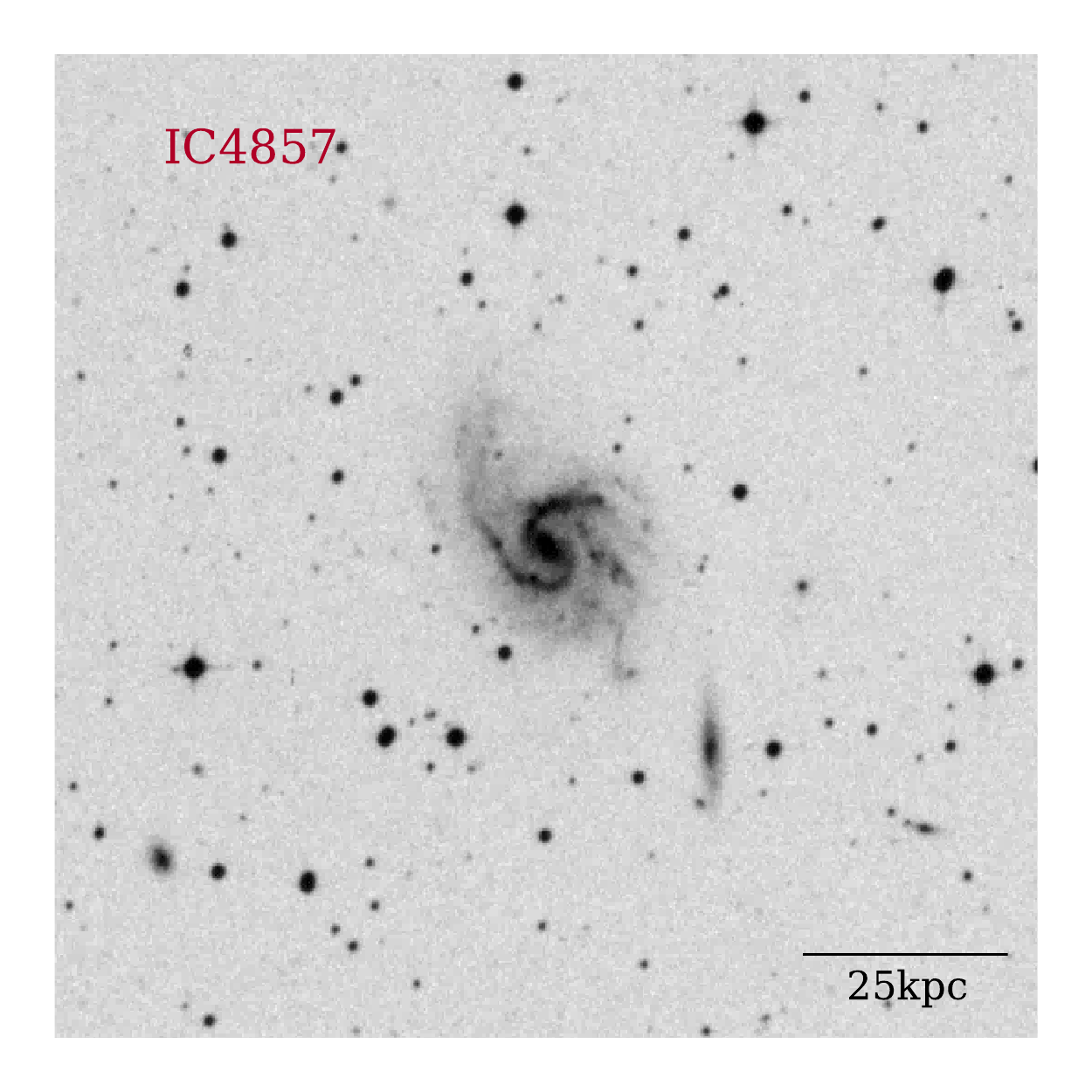} &
\includegraphics[width=0.22\textwidth]{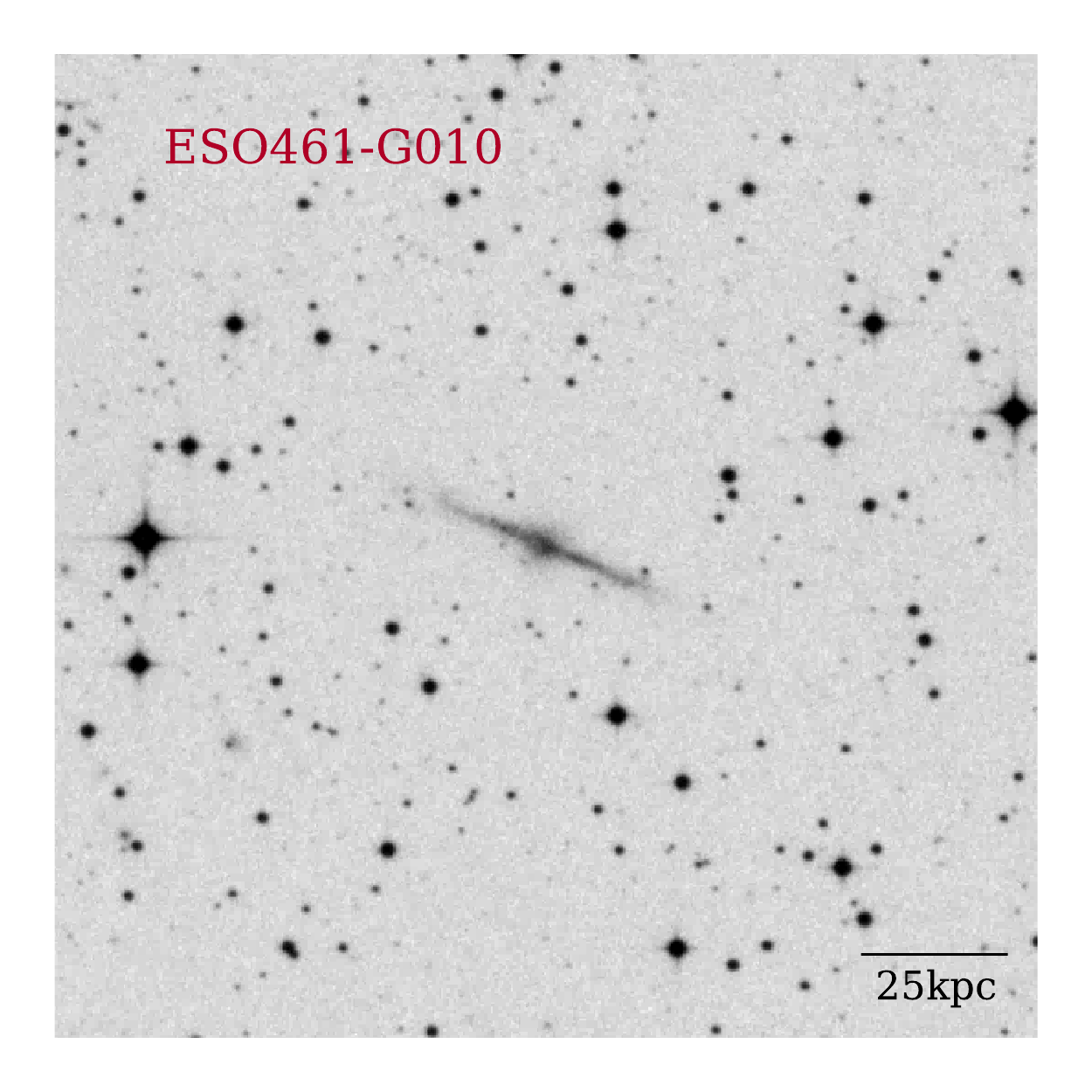} & \includegraphics[width=0.22\textwidth]{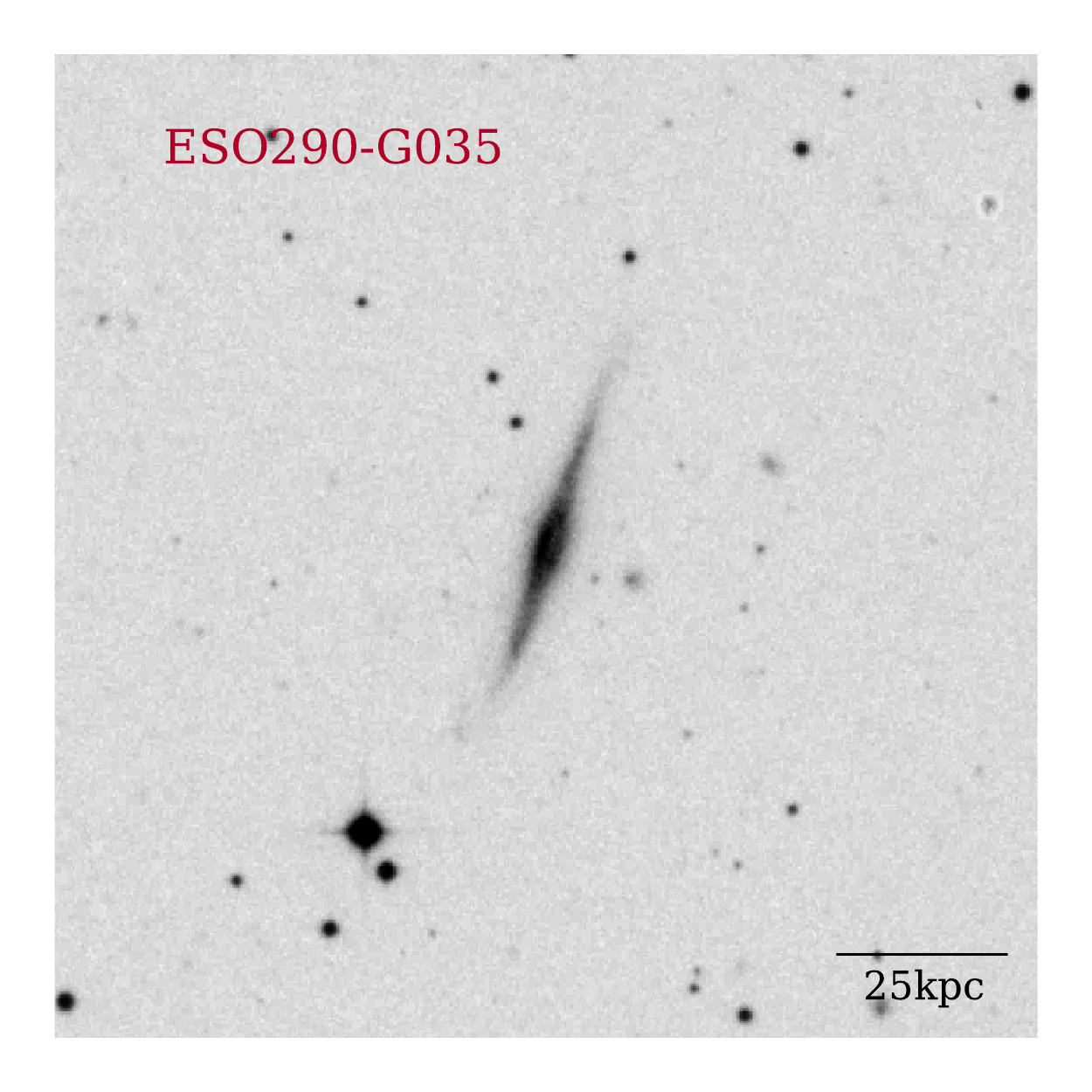} \\
\end{tabular}
\caption{Panel of SuperCOSMOS \bjband\ post stamp images of all \hix\ galaxies. All images are $\rm3\,arcmin$ by $\rm 3\,arcmin$ in size except for the image of NGC\,289, which is $\rm6\,arcmin$ by $\rm 6\,arcmin$ in size. The scale bar in the bottom right corner indicates $\rm 25\,kpc$ at the distance of the galaxy. The optical image of \gal\  is shown further down in Section \ref{sec:focus}.}
\label{fig:optical}
\end{figure*}

\subsection{Estimating galaxy properties}
\label{sec:basic_props}
The \hi\ mass is calculated from the published \hipass\ integrated 21\,cm emission signal by:
\begin{equation}
\label{equ:himass}
\rm M_{HI} [M_{\odot}] = \frac{2.356\cdot 10^5}{1 + z} \cdot (D [Mpc])^2 \cdot F_{HI} [Jy\,km\,s^{-1}]
\end{equation}
with z being the galaxy's redshift, D the distance to the galaxy in Mpc and $\rm F_{HI}$ the integrated flux density in units of Jy\,\kms. The error estimation closely follows the error estimation of $\rm F_{HI}$ as suggested by \citet{Koribalski2004}:
\begin{equation}
\rm S/N = \frac{S_{peak}}{\sqrt{rms^2 + (0.05 \cdot S_{peak})^2}}
\end{equation}
\begin{equation}
\rm \Delta F_{HI}  = \frac{4}{S/N} \cdot \sqrt{S_{peak} \cdot F_{HI} \cdot dv}
\end{equation}
where $\rm S_{peak}$ is the peak intensity in the spectrum, rms the root mean square of the data cube and dv the velocity resolution. Using Gaussian error propagation the error of $\rm F_{HI}$ is then propagated and combined with the error of the distance to estimate the error of the \hi\ mass. 

All stellar masses for the \hix, control and \hipass\ samples were estimated following equation (3) in \citet{Wen2013}, which assumes a linear relationship between the log of the stellar mass and the log of the Galactic extinction corrected \mass\ \kband\ luminosity. 

Baryonic masses for \hix\ and control sample are the sum of atomic and stellar mass. The atomic gas mass is the \hi\ mass increased by 35\,per\,cent to include Helium.

Star formation rates (SFR) for the \hix, control and \hipass\ samples are estimated using the \galex\ \nuv\ luminosity in combination with the \wise\ \wthreeband\ luminosity to account for dust obscured star formation. We use the prescription in equation (1) and (2) in \citet{Saintonge2016} assuming a \citet{Chabrier2003} initial mass function (IMF). Where \galex\ photometry is not available, the SFR prescription by \citet{Cluver2014}, which is based solely on the luminosity in the \wthreeband, is used. Their IMF is consistent with a \citet{Chabrier2003} IMF and the comparison of SFRs estimated with the both methods does not reveal systematic offsets. \gal\ is not detected in the \wthreeband. Its SFR does therefore not include a mid-IR component. 

The gas-phase oxygen abundance is estimated with the O3N2 and N2 method using the H$\alpha$, H$\beta$, [O\,III]$\lambda$\,5007\,\AA\ and [N\,II]$\lambda$\,6584\,\AA\ emission lines as described by \citet{Pettini2004}.

\section{The HIX sample in context}
\label{sec:compare}

In this section the \hix\ galaxy sample is presented and compared to the control sample, the parent sample and other \hi\ rich samples, where appropriate data are available. 

As a first overview, Figure \ref{fig:optical} shows the optical images of all \hix\ galaxies except \gal, for which the optical image is shown Section \ref{sec:focus}. All \hix\ galaxies are spiral galaxies. One of them shows clear signs of interaction (ESO245-G010), the remaining galaxies appear quite regular. 

\subsection{HI mass and stellar mass}
\label{sec:mhi_mstar}
The relation between gas fraction (defined as $\rm M_{HI} / M_{\star}$) and stellar mass is well known and used here to test the \hix\ selection criteria and compare the \hix\ galaxies to the control and parent sample as well as other samples. 

\begin{figure}
\includegraphics[width=3.15in]{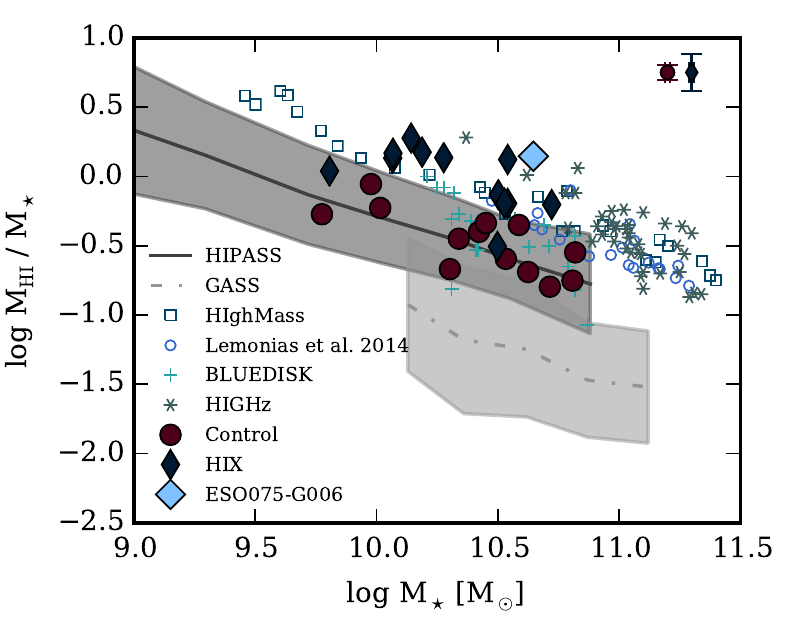}
\caption{The \hi\ gas mass fraction as a function of the stellar mass. The blue diamonds present the \hix\ sample and red circles the control sample. The light blue diamond is \gal, which will be discussed in more detail further below. Blue and red errorbars in the top right corner give the median errorbars for the \hix\ and control sample respectively. The dark grey, solid line presents the running average of the \hipass\ parent sample along with the $1\,\sigma$ scatter as the grey shaded region. The light grey, dashed line and shaded area represents the running average and $1\,\sigma$ scatter of the GASS sample. Open symbols present other surveys of (\hi\ rich) galaxies: \highmass\ (squares), \citet{Lemonias2014} (circles), HIGHz (asterisks) and \bluedisk\ (crosses).}
\label{fig:fhi_vs_mstar}
\end{figure}

Figure \ref{fig:fhi_vs_mstar} shows this relation. Most \hix\ galaxies are located above the $1\,\sigma$ scatter of the parent sample, while the control galaxies populate the \hipass\ scatter. The two \hix\ galaxies that are located within the grey shaded area are ESO208-G026 and IC\,4857. They have been selected to be \hi\ excess galaxies according to the \rband\ scaling relation of \cite{Denes2014}. On the gas mass fraction vs. stellar mass plane, however, they lie within the scatter of the \hipass\ parent sample and thus, their \hi-richness on that scale is not as obvious as for the other \hix\ galaxies. In a future analysis of spatially resolved \hi\ data of these galaxies, we will determine whether these two galaxies are true \hix\ galaxies, more ``normal'' \hipass\ galaxies or something in between.

For comparison, Figure \ref{fig:fhi_vs_mstar} also shows the running average of the GASS sample (data release 3, \citealp{Catinella2010, Catinella2012, Catinella2013}). Non-detections are included as upper limits. GASS is a stellar mass selected sample and as such not biased towards \hi\ rich systems as it is the case for \hipass. The running average of the GASS sample is therefore lower than for the \hipass\ parent sample and even our control sample appears \hi\ rich compared to GASS. This further emphasises that the \hix\ galaxies are among the most \hi\ rich galaxies in the local universe.

We further compare the \hix\ sample to other surveys of \hi\ rich galaxies:  \highmass\ \citep{Huang2014}, a sample by \citet{Lemonias2014}, \hi-Monsters \citep{Lee2014}, HIGHz \citep{Catinella2015} and \bluedisk\ \citep{Wang2013}. For \highmass\ and \citet{Lemonias2014} sample galaxies, SFRs and stellar masses were calculated the same way as for the \hix\ galaxies. For the HIGHz and \bluedisk\ galaxies those values were taken from the respective publications. For all samples, distances and \hi\ masses were adopted from  the respective publications.

\highmass\ and \citet{Lemonias2014} galaxies have been selected from \alfalfa\ \citep{Giovanelli2005} using the $\rm \log f_{HI}$ - $\rm \log M_{\star} [M_{\odot}]$ scaling relation. Thus, both samples are also located above the $1\,\sigma$ scatter of \hipass. While the \citet{Lemonias2014} sample focusses on early type, AGN hosting galaxies, \highmass\ galaxies are spirals. HI-Monsters are HI massive galaxies ($\rm \log M_{HI} [M_{\odot}] > 10.3$) from \alfalfa, i.e. very similar to \highmass. The HI-Monsters sample includes in addition a selection of low surface brightness (LSB) galaxies like Malin\,1. HIGHz are a sample of the highest redshift HI detections of single galaxies. The \bluedisk\ sample has been selected to be \hi\ rich for their morphology and star formation activity. Figure \ref{fig:fhi_vs_mstar} shows that these \bluedisk\ galaxies lie well within the $1\,\sigma$ scatter of \hipass.

In being \hi\ massive and spiral galaxies, the \highmass\ (and \hi-Monsters) sample galaxies appear to be the most similar to the \hix\ sample. 

As none of the \hix\ galaxies is a LSB galaxy and most of the \hi-Monsters have larger stellar masses than the \hix\ galaxies, there is little overlap between these studies. Where both samples overlap, they host comparably massive \hi\ reservoirs at a given stellar mass.

Similar to the \hix\ survey, the \highmass\ survey aims to select the most \hi\ extreme galaxies from an \hi\ selected survey (in this case from \alfalfa). To compare the two sample selections, we use the \mass\ stellar masses for our \hipass\ parent sample. These are only available for 1475 out of 1796 galaxies. 
We further cut this sample down to galaxies, that (a) are located at dec$< -30$\,deg, (b) are more massive than our stellar mass cut ($\rm M_{K} < -22.0$\,mag) and (c) are neither in close proximity to the Galactic plane (for good photometry) nor a massive neighbouring galaxy (exclude source confusion). Applying the \highmass\ sample selection to this parent sample yields 22\,galaxies. These 22\,galaxies include all \hix\ galaxies except for the two galaxies that are also located in the \hipass\ 1\,$\sigma$ scatter in Figure \ref{fig:fhi_vs_mstar} (ESO208-G026, IC\,4857). The remaining 11\,galaxies in this ``\hipass-\highmass'' sample do not appear in the \hix\ survey because they are not outliers to the \citet{Denes2014} scaling relation. This means, when estimating their \hi\ mass from their \rband\ luminosity, the estimated \hi\ mass is not 2.5\,times larger than their measured \hi\ mass. Three of the \highmass\ like galaxies that are not in \hix\ have large stellar masses ($\rm \log M_{\star} [M_{\odot}] \approx 11.0$). There are, however, no differences found, when comparing the distribution of \hi\ mass and \hi\ mass fraction of the \hix\ like galaxies and \highmass\ like galaxies above our stellar mass cut. 

\subsection{Star formation activity}
\label{sec:sfr}

In order to further characterise the sample of \hix\ galaxies, we investigate their star formation activity and compare it to the control sample. 

\begin{figure}
\includegraphics[width=3.15in]{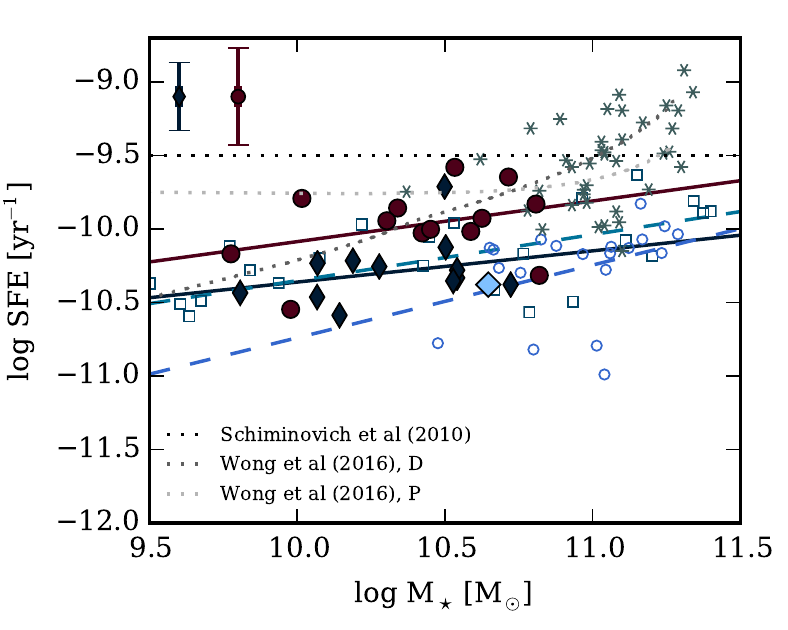}
\caption{The star formation efficiency as a function of stellar mass. Symbols are as in Figure \ref{fig:fhi_vs_mstar}. The red, solid line is a fit to the \hix\ data and the blue, solid line a fit to the control data. The black, dotted line is the average SFE for $\rm log\,M_{\star} / M_{\odot} > 10$ galaxies found by \citet{Schiminovich2010} and the grey, dotted lines are model predictions by \citet{Wong2016}. For comparison, the light blue, dashed lines are fits to the \highmass\ and \citet{Lemonias2014} samples respectively (colours as for their symbols).}
\label{fig:sfr_vs_mstar}
\end{figure}

In Figure \ref{fig:sfr_vs_mstar}, the scatter plot shows the relation between the star formation efficiency (SFE) defined as SFR divided by \hi\ mass, and stellar mass for the \hix\ and control samples. This definition has been used before by e.\,g. \citet{Wong2016} and \citet{Schiminovich2010}. We obtain a linear fits to the data with a Monte Carlo method. This method includes creating 1000 random samples that have the same mean and standard deviation in stellar mass and SFE as the sample, to which the line is fitted. To each of these random samples, a line is then fitted with a standard least squares fit. The final slope and interception of the fit is then mean of the results of the 1000 random samples.

The average specific star formation rates of \hix\ ($\rm log sSFR [yr^{-1}] = (-10.3 \pm 0.2)$) and control sample ($\rm log sSFR [yr^{-1}] = (-10.4 \pm 0.2)$) are comparable, i.\,e. at a given stellar mass both samples form stars at a similar rate. However, as \hix\ galaxies are more \hi\ rich than the control sample, their SFE are systematically lower as can be seen in Figure \ref{fig:sfr_vs_mstar}. 

IC\,4857 is the only \hix\ galaxy that forms stars at similar efficiency as the control sample. ESO123-G023 and ESO240-G011 are the least efficient control galaxies. 

The model by \citet{Wong2016} is calibrated on SINGG galaxies (HI selected, \citealp{Meurer2006}) and comes in two flavours: The relative contributions by \hi\ and \htwo\ are disentangled using a prescription either dependent on the stellar surface mass density (D) or the hydrostatic pressure (P). While the hydrostatic pressure model is the preferred model by \citet{Wong2016}, the stellar surface mass density model better describes the SFEs of the control sample. The \hix\ galaxies form stars less efficiently than suggested by both models. All galaxies form stars less efficiently than the average value found by \citet{Schiminovich2010}. It is to be noted though that the scatter of their data is three orders of magnitudes. The SFEs of all \hix\ and control galaxies are larger than $\rm 10^{-10.75}\,yr^{-1}$. Below this SFE, \citet{Schiminovich2010} consider a galaxy passive for their \hi\ reservoir. 

Linear fits to the SFEs of the \highmass\ and \citet{Lemonias2014} galaxies are also shown in Figure \ref{fig:sfr_vs_mstar}. In particular the linear fit to the \highmass\ galaxies is very close to that of the \hix\ sample further emphasising the similarity of the two samples. \citet{Lemonias2014} galaxies tend to be less efficient than the \hix\ galaxies. The HIGHz sample is the most efficient at forming stars from their massive \hi\ reservoirs. 

To summarise, \hix\ galaxies form stars  less efficiently than the control sample. As outlined in the introduction, one reason for decreased SFE can be the increased angular momentum of a galaxy. In the following section we will discuss \gal, for which the resolved ATCA \hi\ data will be analysed and the increased angular momentum hypothesis will be tested.

\section{The most extreme: \gal}
\label{sec:focus}

\gal\ hosts the most massive \hi\ disc of the \hix\ galaxies with an \hi\ mass of log M$\rm _{HI}$  [M$_{\odot}$] = $(10.8 \pm 0.1)$ as measured in \hipass. This measured \hi\ mass is 2.6 times larger than expected from its \hopcat\ \rband\ magnitude. An \hi\ mass of almost $\rm 10^{11}\,M_{\odot}$ is rare. According to the \hi\ mass function by \citet{Martin2010} based on \alfalfa, in a volume of 400,000\,Mpc$^{3}$ one expects to find only one galaxy to hosts an as massive \hi\ disc as \gal. 

In this section the optical morphology, the \hi\ properties and kinematics, the optical spectra and the local environment of \gal\ will be discussed. This section aims to understand the reason for the \hi\ excess in the most extreme \hix\ sample galaxy. Similar analysis for the remaining \hix\ galaxies and a comparison to the control sample will be presented in a future paper.

\subsection{Optical morphology of \gal}
The SuperCOSMOS \bjband\ image of \gal\ is shown in panel (a) of Figure \ref{fig:mom_ana}. \gal\ hosts a bar and seems to be accompanied by a dwarf galaxy located to the north-west. In the optical an inner ring and an outer pseudo ring is visible as the RC3 morphological classification indicates. The inner ring is also visible in \galex\ \nuv\ images as a star forming ring (see grey scale image in Figure \ref{fig:metal}, exposure time: 207\,s). As both the SuperCOSMOS and \galex\ \nuv\ imaging is relatively shallow, faint spiral arms or an XUV disc can not be excluded. There is no \galex\ \fuv\ image available. The stellar mass is calculated as $\rm \log M_{\star} [M_{\odot}] = 10.5$ and the ESO-LV catalogue gives a 25\magasec\ isophotal radius of $\rm30.2\,arcsec = 21.7\,kpc$. 

As some of the most \hi\ massive galaxies known, e.\,g. Malin 1 \citep{Pickering1997}, are low surface brightness (LSB) galaxies, \gal\ is a good candidate for a LSB galaxy. However, when using the central \kband\ surface brightness as indicator \citep{MonnierRagaigne2003}, the value of 17.41\magasec\ of \gal\ is brighter than the threshold to LSB discs. 

\subsection{HI morphology and radial column density profiles}
The two  \hi\ data cubes for \gal\ have been produced as explained in Section \ref{sec:obs-data}. In this section we use the robust=0.5, dv=4\kms\ data cube to analyse the distribution of \hi\ in \gal. 

\begin{figure}
\includegraphics[width=3.15in]{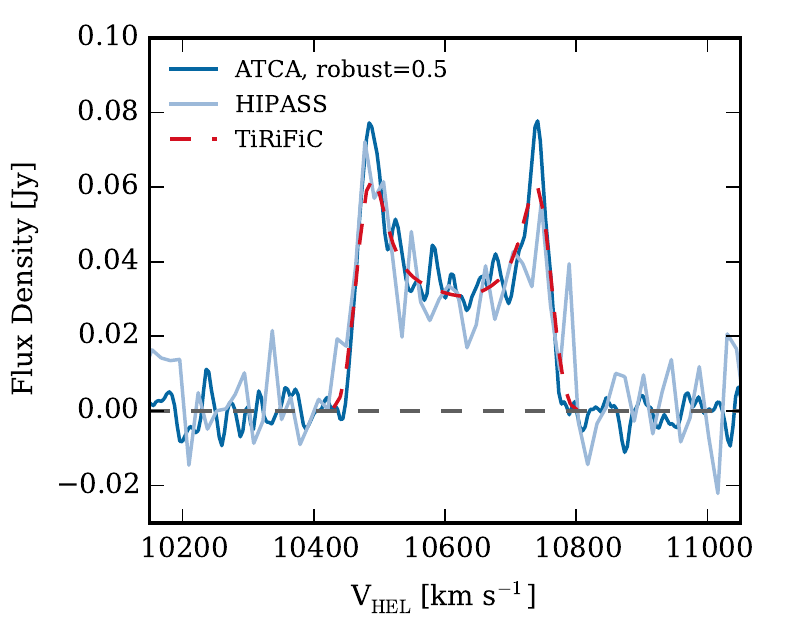}
\caption{Comparison of \hi\ spectra as measured from the \hipass\ data cube (light blue), the ATCA data cubes (dark blue for robust=0 and medium blue for robust=0.5) and the \tirific\ model data cube (red dashed line, see Section \ref{sec:tirific}).}
\label{fig:spec}
\end{figure}

Figure \ref{fig:spec} shows the spectrum as taken from the ATCA data cube and from the \hipass\ data. They agree within the noise level. The double horn is close to symmetric indicating that the approaching a receding half of the disc contain approximately the same amount of gas. The integrated flux density of \gal\ measured from the ATCA data amounts to $\rm 12.8\,Jy\,km\,s^{-1}$. Using Equation \ref{equ:himass} this translates into $\rm \log\,M_{HI} [M_{\odot}] = 10.8 \pm 0.1$, which is in agreement with the \hipass\ measurement of $\rm \log\,M_{HI} [M_{\odot}] = 10.8 \pm 0.2$ and indicates that the ATCA data cube is not missing any diffuse \hi. 

\begin{figure*}
\begin{tabular}[H]{c c}
\includegraphics[width=3.15in]{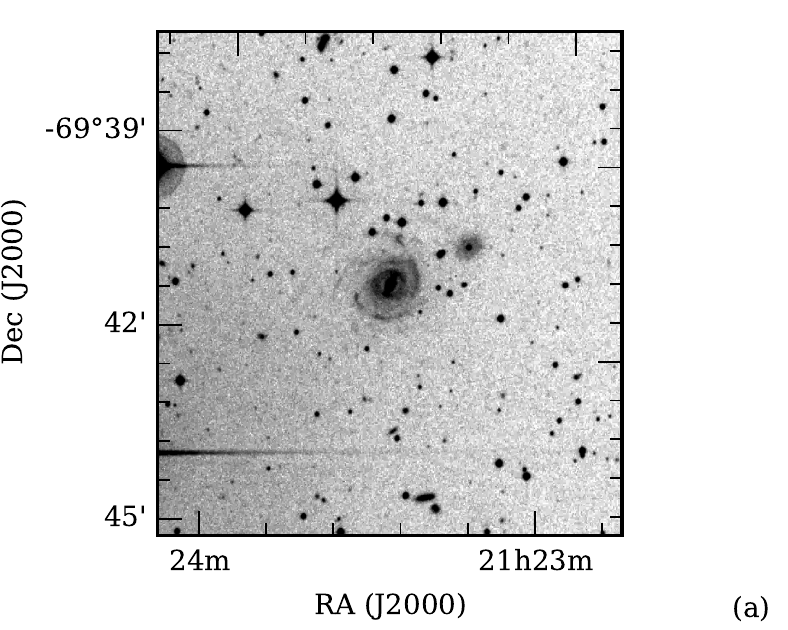}  & \includegraphics[width=3.15in]{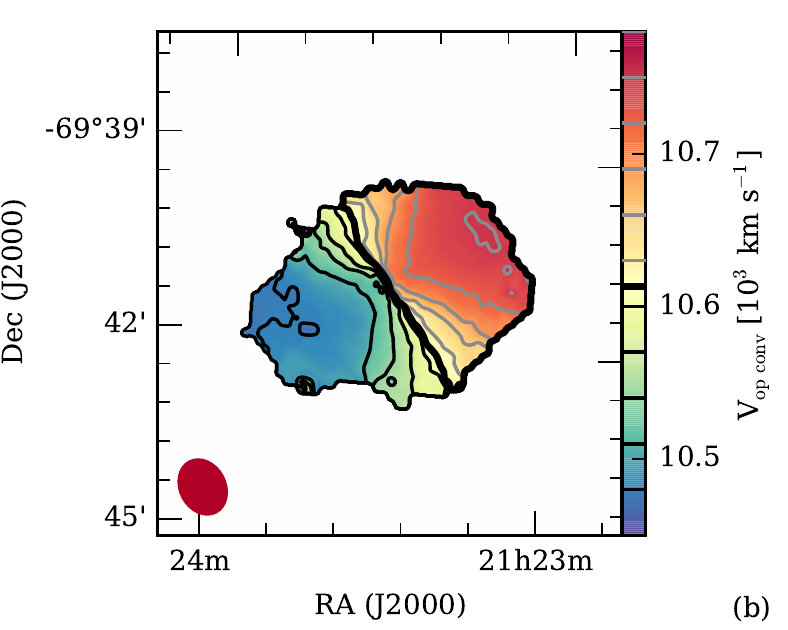} \\
\includegraphics[width=3.15in]{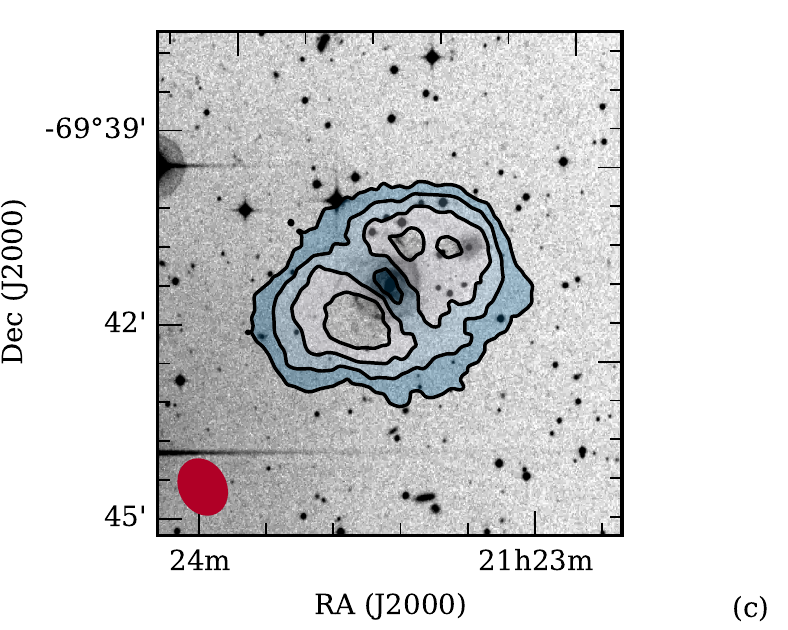} & \includegraphics[width=3.15in]{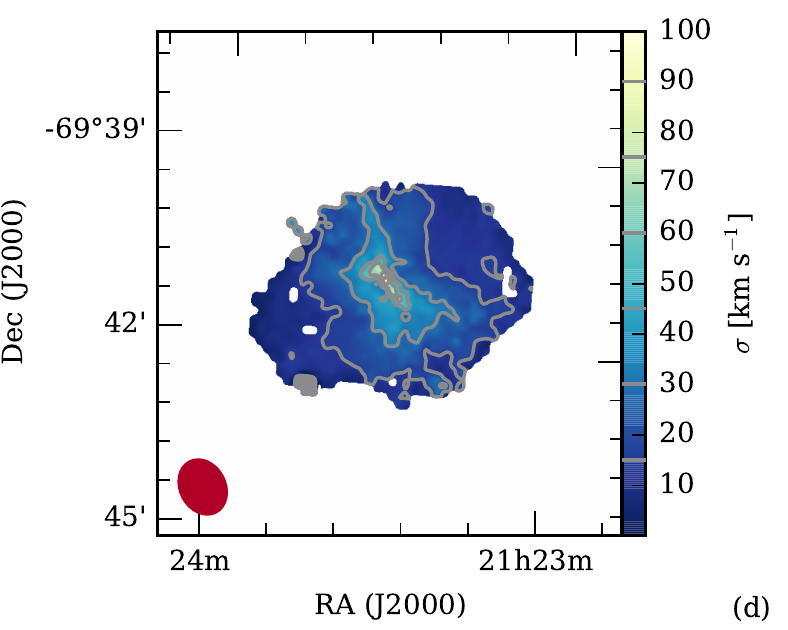} \\
\end{tabular}
\caption{\textbf{(a)} The optical SuperCOSMOS \bjband\ image of \gal. \textbf{(b)} The velocity field (moment 1 map) of \gal\, solid and dashed contours are separated by 20\kms and indicate the receding and approaching side respectively. The thick black contour is drawn at 10613\kms indicating the systemic velocity. \textbf{(c)} The column density (moment 0) map of ESO075-G006 overlaid on the SuperCOSMOS optical B-band image. \hi\ column density contours are at (0.3, 1.3, 2.3, 3.3)\msunpcsq. \textbf{(d)} The velocity dispersion (moment 2) map of \gal. Contours are separated by 15\kms. The red ellipses in the bottom left corners of the moment maps indicate the synthesised beam size. All moment maps are created from the rob=0.5 cube.}
\label{fig:mom_ana}
\end{figure*}

The \hi\ column density contours overlaid on the optical SuperCOSMOS image are shown in Figure \ref{fig:mom_ana}\,(c). The locations of maximum column density in the \hi\ disc are found to the north and south of the centre and align with the ridge of the central bar. Furthermore, they coincide with the outermost spiral arms as visible in \nuv\ and faintly visible in the optical. The approaching and receding half appear very similar, i.\,e. the disc is symmetric. No obvious signs of interaction and tails are visible. In the centre of the disc the \hi\ column density drops to approximately 1\msunpcsq. This is a common feature of \hi\ discs and hints to molecular gas in the centre of \gal\ \citep{Bigiel2012}. The distance between the geometrical centre of the \hi\ disc and the 2MASX coordinates of the stellar disc centre is less than $\rm 10\,arcsec$, which is smaller than the spatial resolution element. Therefore, both coincide. In conclusion, the \hi\ morphology points towards a well settled disc.

\begin{figure}
\includegraphics[width=3.15in]{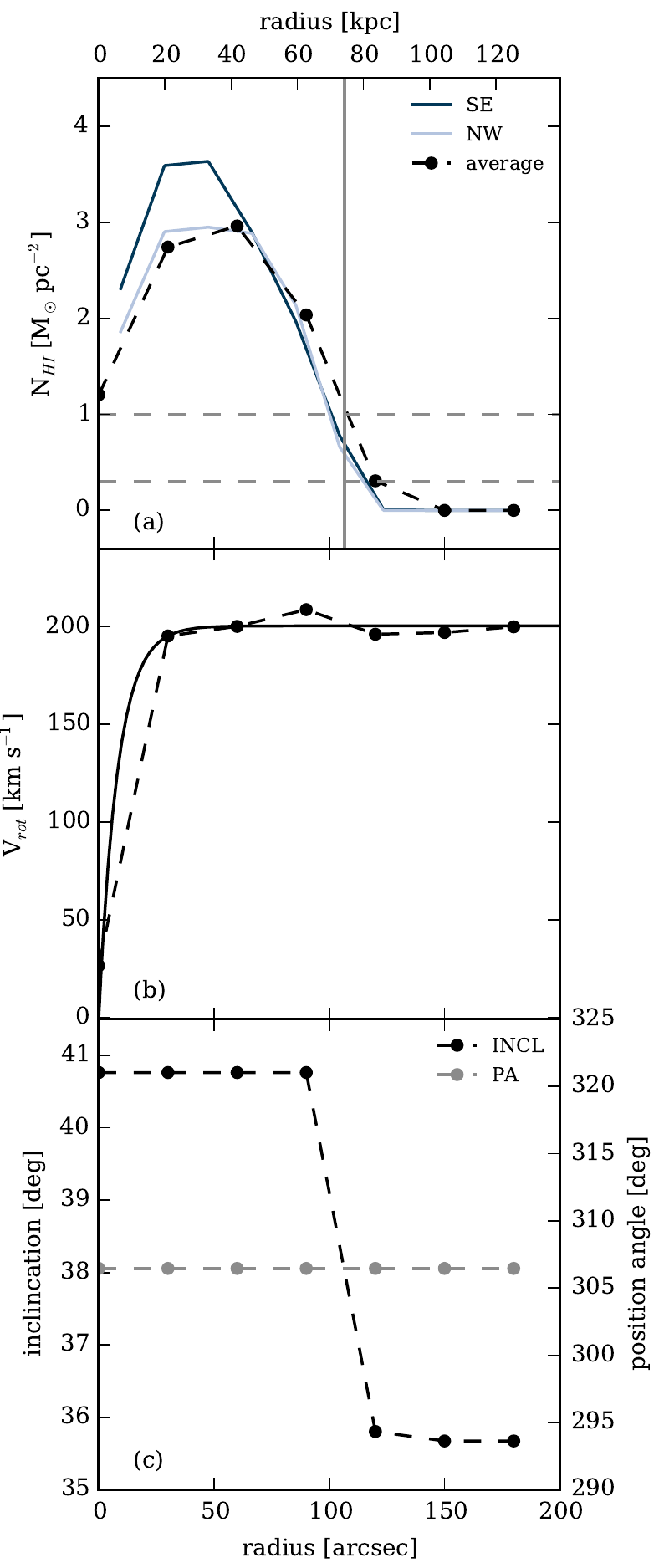}
\caption{Radial profiles of HI properties: \textbf{(a):} The radial \hi\ column density profile of \gal\ in azimuthally averaged rings (black points and dashed line) and along the major axis from the centre going north-westerly (south-easterly) in light (dark) blue lines. The bottom grey dashed line marks the lower limit on column densities of $\rm 0.4 \cdot 10^{20}\,cm^{-2}$, the top grey dashed line marks the 1\msunpcsq\ level. The vertical grey, solid line marks the measured $\rm R_{HI}$.  \textbf{(b):} Rotation curve from \tirific, the black dashed line connects the measurements and the black solid line is a fit to the data. \textbf{(c):} The radial variation of the inclination (black) and the position angle (grey) as fitted by \tirific.}
\label{fig:hi_radial_profile}
\end{figure}

Radial profiles of the \hi\ column density are obtained from the moment 0 map by measuring the median column density in co-centric elliptical annuli. In Figure \ref{fig:hi_radial_profile}\,(a) an azimuthally averaged profile is shown. To account for possible disc asymmetries, the annuli were also divided in four quadrants with opening angles of 90\,deg centred on either a semi-major or semi-minor axis. This allows for profiles along the semi-major axis in north-western and south-eastern direction. The two profiles along the major axis are also presented in Figure \ref{fig:hi_radial_profile}\,(a). Within the errors of the measurements, the approaching and receding sides show the same radial column density behaviour, again pointing towards a regular \hi\ disc. 

The size of an \hi\ disc is commonly quantified through $\rm R_{HI}$, which is the radius at which the column density of the \hi\ drops to 1\msunpcsq. \citet{Broeils1997} and later \citet{Wang2014,Wang2016} have found  $\rm R_{HI}$ to correlate remarkably well with the \hi\ mass of the disc. This relation indicates that the average \hi\ column density in settled \hi\ discs is constant from galaxy to galaxy. For \gal\ $\rm R_{HI} = (74 \pm 4)\,kpc$ is measured from the azimuthally averaged radial profile in Figure \ref{fig:hi_radial_profile}\,(a) and corrected for beam smearing following \citet{Wang2014,Wang2016}. The measured value is then compared to the expected $\rm R_{HI, exp}$ that is calculated from equation (2) in \citet{Wang2014,Wang2016}. We find  $\rm R_{HI, exp} = (74 \pm 18)\,kpc$. The measured and expected value are in agreement within the errors. Hence, the average \hi\ column density of \gal\ is normal. 

\subsection{HI kinematics}
In this section we investigate the \hi\ kinematics in detail searching for evidence of recent gas accretion e.\,g. radial motions or vertical gradients indicating a galactic fountain. 
 
\subsubsection{2D moment analysis}
 
Panel (b) of Figure \ref{fig:mom_ana} shows the velocity field of \gal. The iso-velocity contours show the typical spider-web pattern of regularly rotating discs. Radial motions can induce S-shaped iso-velocity contours in the central parts of the galaxy. This, however, is not seen and is a first indication that radial motion are not playing a major role in the velocity field of \gal. 

Figure \ref{fig:mom_ana}\,(d) shows the velocity dispersion of the \hi\ disc of \gal. The low velocity dispersion in the outer parts, especially along the major axis of the disc further indicates a regularly rotating disc. The increase of the measured velocity dispersion towards the centre of the disc is due to projection effects. 

\subsubsection{3D kinematic analysis}
\label{sec:tirific}

To further explore the kinematic properties of \gal, a tilted ring model is fitted to the robust=0, dv=10\kms\ \hi\ data cube using \tirific\ \citep{Jozsa2007}. For the fit 7\,rings, each 30\,arcsec wide, are defined. In addition two half discs are defined, each centred on the major axis with an opening angle of 180\,degrees. This way the approaching and the receding side of the \hi\ disc can be fitted independently. However, only the surface brightness is allowed to vary between the half discs, all other quantities are fitted to full rings. A variety of models were fitted to the data:
\begin{enumerate}
 \item a flat rotating disc.
 \item a warped disc with purely rotational motions.
 \item discs with lagging thick discs, i.\,e. Galactic Fountain mechanism.
 \item flat and warped discs with radially constant or varying radial velocity components.
 \item rotating discs with a combination of radial velocity components and vertical velocity gradients.
\end{enumerate}

All models return similar values for surface brightness, systemic and rotation velocity, inclination (range), position angle (range) and kinematic centre. The flat disc model is not able to produce a flat rotation curve at large radii but requires a sharp drop in the rotation velocity by 30\kms. This model is therefore excluded. More complicated and exotic models (models (iii), (iv) and (v)) do not trace emission in channel maps and position--velocity diagrams better than a warped disc with purely rotational motions. The data are limited by sensitivity and spatial resolution. The effects investigated in models (iii), (iv) and (v) can still occur in \gal\ but we are not able to detect evidence in favour of them. We apply Occam's Razor and select the simplest model with a flat rotation curve, which is the warped disc. In the following, this model is presented and discussed.

\begin{figure*}
\includegraphics[width=160mm]{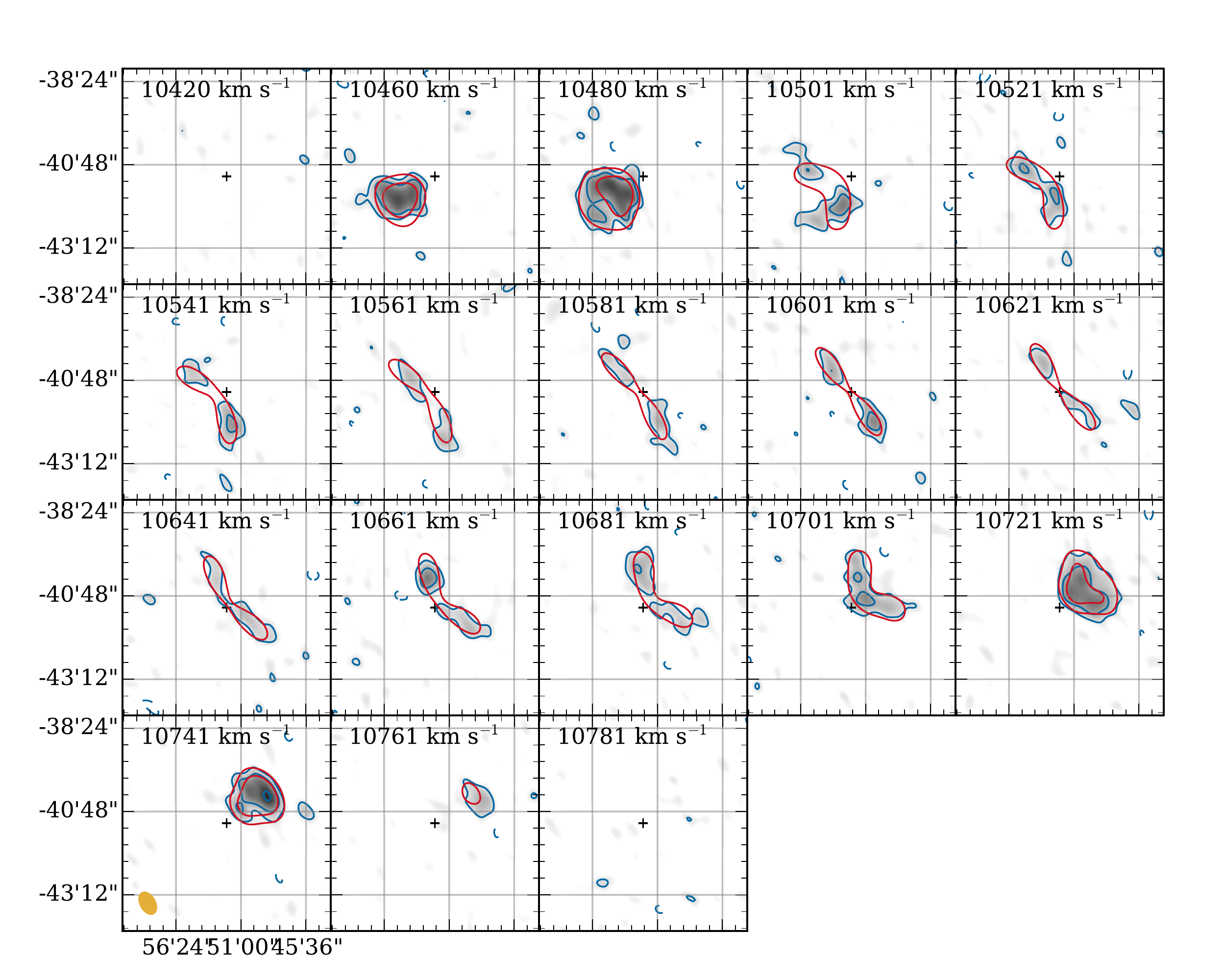}
\caption{A selection of single channel maps from the \hi\ data cube of \gal\ (blue contours) overlaid on the grey scale image of the channel. The red contours show the \tirific\ fit to the data. Both sets of contours are  (-3.6, 3.6, 7.2, 14.4)$\rm\,mJy\,beam^{-1}$. The velocity of each channel is given in the upper left corner. The black cross marks the centre of the galaxy. The coordinates are the same in all channels. For simplicity only minutes and seconds in the first panel are given.}
\label{fig:channel_map}
\end{figure*}

The rotation curve is shown in Figure \ref{fig:hi_radial_profile}\,(b). Fitting a rotation curve of the functional form:
\begin{equation}
\rm v_{rot} (r) = v_{flat} \cdot \left[ 1 - exp\left(\frac{-r}{l_{flat}}\right) \right]
\end{equation}
(\citealp{Leroy2008} and references therein) yields a rotation velocity of $\rm v_{flat} = (201 \pm 6)\,km\,s^{-1}$. 

The model suggests a warp in inclination, which is dropping from $\rm 41\ to \ 36\,deg$  between radii of 90 and $\rm 120\,arcsec$. The radial variation of inclination, i.\,e. the warp is shown in Figure \ref{fig:hi_radial_profile}\,(c). This is a warp at the lower end of the warp scale found e.\,g. in the THINGS galaxies \citep{Schmidt2016}. The position angle is held constant with radius and is fit to 306\,deg (see again Figure \ref{fig:hi_radial_profile}\,(c)).   

Furthermore, \tirific\ finds a velocity dispersion $\sigma$ of 11.2\kms and a systemic velocity of 10613\kms. The kinematic centre of the disc is located at $\rm 21h\,23\arcmin\,29.5\arcsec\ {-69\degr}\,41\arcmin\,06\arcsec$, which is the same location as the 2MASX centre of the stellar disc. 

Figure \ref{fig:channel_map} shows selected channel maps of the input data and resulting model cube. The model performs especially well in the velocity range between $\rm 10521\ and\ 10721$\kms. More elaborate models have not been able to decrease the residuals in channels 10501\kms. 

\subsection{Stability of \gal's disc}

\begin{figure}
\includegraphics[width=3.15in]{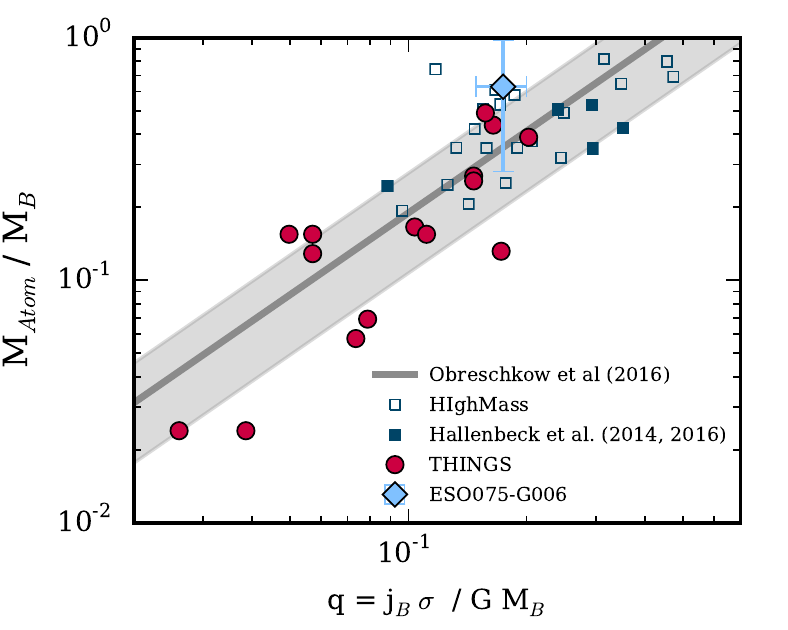}
\caption{The atomic to baryonic mass fraction as a function of the global stability parameter q. The blue diamond presents our measured data for \gal. The red circles present the data of THINGS galaxies \citep{Walter2008} as measured by \citet{Obreschkow2014}. The black line and grey shaded area mark the analytical model by \citet{Obreschkow2016}. Blue squares present \highmass\ galaxies. Filled square indicate galaxies for which resolved \hi\ maps have been published \citep{Hallenbeck2014,Hallenbeck2016}.}
\label{fig:stability}
\end{figure}

Combining the results of the \tirific\ fit with radial profiles of \hi\ and stellar mass, a global stability parameter can be determined for \gal\ following \citet{Obreschkow2014, Obreschkow2016}. 

\hi\ and stellar masses are calculated in elliptical concentric annuli. Inclination and position angle for each annulus are taken from the \tirific\ result. For high sensitivity, the \hi\ mass map is measured from a moment 0 map created from the robust=0.5 data cube without clipping. For the stellar masses, luminosities are measured from \mass\ \kband\ images and converted to stellar masses again following equation (3) in \citet{Wen2013}. Together with the \tirific\ measurement of the rotation curve, the integrated specific baryonic angular momentum calculated by:
\begin{equation}
\label{equ:jb}
\rm j_{B} = \frac{\sum_{i} (M_{HI, i} + M_{\star, i}) \cdot V_{rot, i} \cdot r_{i}} {\sum_{i} (M_{HI, i} + M_{\star, i}) }.
\end{equation}
$\rm M_{HI, i}$ and  $\rm M_{\star, i}$ are the \hi\ and stellar mass and $\rm V_{rot, i}$ the rotation velocity in the i$\rm ^{th}$ annulus with radius $\rm r_i$. Equation \ref{equ:jb} assumes that stars and \hi\ rotate at the same speed.

The global stability parameter is then calculated by:
\begin{equation}
\rm q = \frac{j_B \cdot \sigma}{G \cdot M_B} = \frac{8922\,km\,s^{-1}\,kpc \cdot 11.2\,km\,s^{-1}}{G \cdot 10^{11.2\pm0.1}\,M_{\odot}}
\end{equation}
where G is the gravitational constant. We find $q = 0.16 \pm 0.02$. In Figure \ref{fig:stability} the atomic to baryonic mass fraction is shown as a function of q. The data point for \gal\ agrees with the analytic model by \citet{Obreschkow2016}. This model assumes a flat, exponential disc, where gas is converted into stars when it is not Toomre-stable \citep{Toomre1964} and gas is stabilised by the specific baryonic angular momentum of the disc. For \gal\ this implies, that the star formation efficiency is decreased by the relatively high specific baryonic angular momentum of the disc. The galaxy is therefore able to build up and support a very massive \hi\ disc. It is still an \hi\ excess galaxy for its optical luminosity. For its specific baryonic angular momentum, however, it hosts a normal amount of \hi.

The angular momentum of the molecular gas was not included in this analysis, as no measurements are available. We argue, however, that the molecular gas does not contribute significantly to the baryonic angular momentum as most of the molecular gas will be located at small radii and the estimated molecular gas mass is smaller than both the \hi\ and the stellar mass.

For comparison, we also show the data for THINGS galaxies \citep{Walter2008,Obreschkow2014,Obreschkow2016}, the only sample of massive galaxies for which this detailed analysis has been performed. We furthermore estimate the stability parameter for those \highmass\ galaxies, for which a rotation velocity and a exponential scale length or a 25\,\magasec\ isophotal radius has been published. For this estimate the stellar specific angular momentum is calculated following equations (4) and (6) in \citet{Obreschkow2014}. The specific angular momentum of the atomic gas is assumed to be twice the stellar specific angular momentum. For the baryonic specific angular momentum, these estimates are combined using Equation \ref{equ:jb}. Except for AGC\,248881, the \highmass\ galaxies follow the model. The outlier is located in close proximity to other massive galaxies and its kinematics might therefore be altered. Encouraged by the success of this model to explain the extremely high \hi\ mass of \gal\ and \highmass\ galaxies, we will further populate this graph and test the model with the remaining \hix\ and control galaxies in the future. 

\subsection{The local environment of \gal}

The data cube has been searched for dwarf galaxies and remnants thereof. A candidate for such a dwarf galaxy is the object to the north-west of the \gal. \citet{Arp1987} have classified this galaxy together with \gal\ as an interacting double.

Following the Hyper-Leda convention the companion is identified as PGC\,277784 \citep{Paturel2003} and is classified as a non-star object in the Guide Star Catalogue (GSC, \citealp{Lasker2008}). The coordinates of this object are: $\rm 21h\,23\arcmin\,16.0\arcsec\ {-69}\degr\,40\arcmin\,25\arcsec$. Assuming this companion is located at the same distance as \gal, a stellar mass can be estimated from the \mass\ \kband\ magnitude (taken from the \mass\ point source catalogue, \citealp{Cutri2003}). This yields $\rm log\,M_{\star} [M_{\odot}] = 9.0$.

Including the stellar mass of the dwarf in the stability analysis in the previous section does not change the q parameter significantly. To do so, we assumed that the dwarf would travel at the same rotation velocity as the surrounding \hi.  

No separate \hi\ content is detected for PGC\,277784 above the column density limit within the data cube. Assuming this dwarf to be an unresolved point source at the three sigma column density limit and to have a velocity width of 50\kms, the upper limit for its \hi\ mass is $\rm log\ M_{HI} / M_{\odot} \le 8.7 $. With the estimated stellar mass this is equivalent to $\rm log\,f_{HI} = -0.5$, which is well below the $1\,\sigma$ scatter of the \hipass\ running average in Figure \ref{fig:fhi_vs_mstar}.  

The residual data cube of the \tirific\ fit, i.e. the difference between the input data cube and the output \tirific\ model cube, does not reveal excess emission in the area of PGC\,277784. Hence, there is no \hi\ emission within the data cube that can not be explained with the warped \hi\ disc model. It is to be noted though that the spatial resolution is limited and could be too coarse to detect signs of the companion. 

No signature of other galaxies are found within an \hi\ data cube of 1\,deg in side-length and $\rm V_{sys}\pm 600$\kms. 

The nearest, similarly sized neighbour of \gal\ is $\rm 2MASX\,J21200298-6924152$, which is at a larger projected distance than 1\,Mpc and its recession velocity differs by 600\kms. \citet{Crook2007} does not place \gal\ in a group. In summary, \gal\ appears as an isolated field galaxy.

\subsection{Optical spectra of ESO075-G006}

We observed \gal\ with three different \wifes\ pointings along the ridge of the bar. For each of these pointings one stellar population model has been fitted to the median stacked spectra of all spaxels in the pointing, the recession velocity has been measured from the redshift of the H\,$\alpha$ line and gas phase oxygen abundances have been measured for all star forming regions, in which the necessary lines for metallicity estimation have been detected. 

Using the results of the stellar population modelling, a recession velocity of the stellar disc at each pointing has been measured. Furthermore, the redshift of the H$\alpha$ line in all Voronoi bins indicates the rotation of the ionised gas component. All three components, i.\,e. stars, atomic and ionised gas are co-rotating. This indicates that the disc of \gal\ has not been recently (less than one orbital time) disturbed by any major merger. 

\begin{figure}
\includegraphics[width=3.15in]{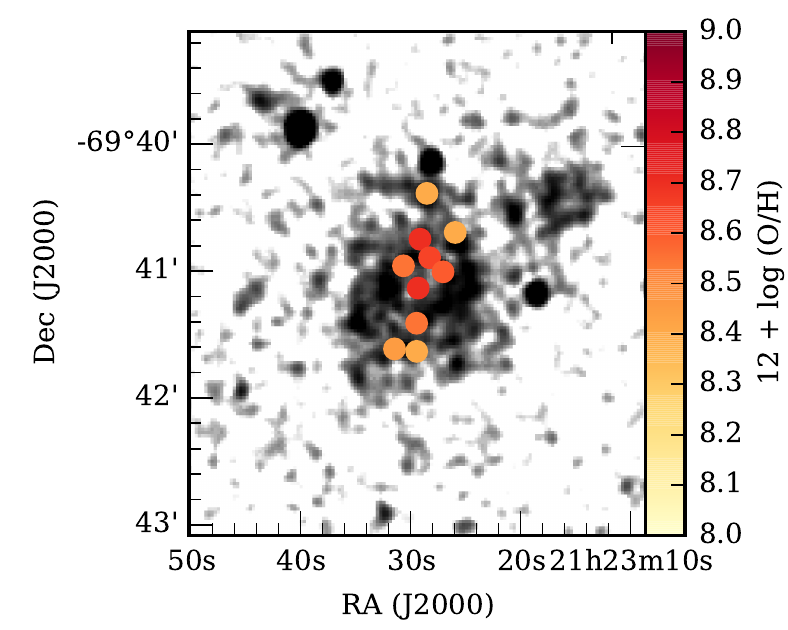}
\caption{The gas-phase oxygen abundance distribution in \gal: The grey shaded image in the background is the \galex\ \nuv\ image. The coloured circles indicated star forming regions within which a metallicity has been measured. The colour of the circle indicates the measure metallicity. }
\label{fig:metal}
\end{figure}

From the emission lines in the \wifes\ data cubes metallicities can be measured as explained in Section \ref{sec:basic_props}. In Figure \ref{fig:metal} those observed (star forming) regions, in which the four necessary emission lines have been detected, are circled. For these region a gas-phase oxygen abundance has been estimated. 

The central metal abundance in the bar amounts to $\rm 12 + log(O/H) = 8.7$. For a comparison we take the Sloan Digital Sky Survey data release 7 (SDSS, DR7, \citealp{Abazajian2009}) in a similar redshift range as the \hix\ galaxies. The stellar mass for these galaxies were taken from the MPA-JHU catalogue\footnote{http://www.mpa-garching.mpg.de/SDSS/DR7/}. From these data we select all those galaxies that lie within $\rm 10.4 \le log\ M_{\star} [M_{\odot}] \le 10.6$ and estimate the metallicity the same way as for \gal\ from emission lines. The average and standard deviation of all estimated metallicities is $\rm 12 + log(O/H) = 8.7 \pm 0.1$. Hence, the central gas--phase oxygen abundance of \gal\ is average for its stellar mass.

In the four regions located on the outer spiral arms / the outer ring, the metallicity amounts to $\rm 12 + log(O/H) = 8.4 \pm 0.1$. \citet{Moran2012} investigated the radial profiles of gas-phase oxygen abundance using the same metallicity estimator. In the same stellar mass range as \gal, they find the metallicity in some galaxies to drop as low as in \gal\ but most galaxies are more metal rich in the outskirts (see figure 4 of \citet{Moran2012}). They further find a relation between the \hi\ mass fraction and the metallicity in the outskirts of a galaxy (see figure 8 of \citet{Moran2012}). For $\rm log\ f_{HI} > 0.0$, they observe the metallicity in the outskirts to drop to values around 8.3, which is similarly metal poor as the outskirts of \gal. This result, however should be taken with caution as the relation between outer metallicity and $\rm log\ f_{HI} > 0.0$ is only sparsely sampled in \citet{Moran2012} and no reliable measurement of $\rm r_{90}$ is available for \gal. Hence, the location of outskirts in \gal\ might be different than in the \citet{Moran2012} galaxies. 

\section{Discussion}
\label{sec:discuss}

In the following, the results from Section \ref{sec:compare} and \ref{sec:focus} will be discussed. We will start with the discussion of \gal\, then continue on to setting the \hix\ galaxies into context.

\subsection{\gal}

The aim of the detailed examination of \gal\ has been to investigate the origin of the \hi\ excess. As indicated in the introduction we are considering \hi\ excess to due both clumpy and smooth accretion as well as due to inefficient star formation. In our data set we can search for evidence for any of following processes: 
\begin{enumerate}
\item non-circular gas motions due to inflowing gas or a lag of the thick disc as produced by a Galactic Fountain. 
\item gas-phase metallicity gradients or inhomogeneities as gas accreted from the intergalactic medium is metal-poor. Metallicity inhomogeneities might coincide with star forming regions, which were triggered by the accretion event \citep{Ceverino2016, Filho2013}.
\item lopsided \hi\ morphologies as filaments of cold gas accretion are distributed randomly.
\item signs of recent mergers, which would disturb both the \hi\ and the stellar disc.
\item a high angular momentum or other factors that reduce the star formation efficiency \citep{Obreschkow2014,Leroy2008}.
\end{enumerate}

The hypothesis that \gal\ went through a recent gas-rich major merger can be dismissed, because both the morphology and the velocity field are very regular and the stellar, atomic and ionised gas components are co-rotating. 

Furthermore, \hi\ richness purely due to a gas-rich minor merger with the dwarf companion in the north-west is also disfavoured. We find no evidence of this dwarf hosting its own distinct gas reservoir. The upper limit for the \hi\ mass of the dwarf is $\rm log\ M_{HI} [M_{\odot}] \le 8.7$, the difference between the measured and expected \hi\ mass of \gal\ is $\rm log\ M_{HI} [M_{\odot}] \le 10.6$. Hence, the detection upper limit in \hi\ mass of the dwarf is two orders of magnitudes too small to explain \gal's \hi\ excess. 

If the \hi\ reservoir of PGC\,277784 has already been incorporated into the \hi\ disc of \gal, we can estimate the amount of \hi\ brought in by PGC\,277784 from its stellar mass and the gas mass fraction - stellar mass relation (see Figure \ref{fig:fhi_vs_mstar}). If PGC\,277784 used to be an average \hipass\ galaxy, this would result in $\rm \log f_{HI} = 0.3$, which is equivalent to an \hi\ mass of $\rm log\,M_{HI} / M_{\odot} = 9.3$. This is still more than an order of magnitude too small to explain the entire \hi\ excess mass. Hence, there needs to be an additional reason for the \hi\ excess of \gal. It is possible, however, that interaction with PGC\,277784 is responsible for the small warp, as the warp occurs at similar galactocentric radii as the location of PGC\,277784.

Our best-fitting \tirific\ model does not need any radial gas motions or vertical velocity gradients to explain the velocity field of \gal. This implies that accretion or a Galactic Fountain could only operate at smaller column densities or spatial scales than can be detected in the \hi\ data of \gal. Hence, \hi\ excess due to a recent phase of violent cosmological accretion as well as a very active Galactic fountain can not be confirmed. We find, however, that the gas--phase oxygen abundance drops in the outskirts of \gal. In general, metallicity gradients are attributed to accretion of pristine, cosmological gas (e.\,g. \citealp{Moran2012}). \citet{Stevens2016i} find in the hydrodynamical, cosmological \textsc{Eagle} simulations that hot halo gas already takes on the structure of the disc before it is accreted on the disc. If hot halo gas accretion is the dominant accretion channel in \gal, it would therefore be near impossible to detect signs of accretion in the morphology and kinematics of the galactic disc. One possible scenario might therefore be that \gal\ is accreting gas but at low column densities or from the hot halo rather than through filaments. 

An explanation for the massive \hi\ disc of \gal\ is provided by the global stability parameter \citep{Obreschkow2016}. Due to its high specific baryonic angular momentum, \gal\ is able to stabilise the gas against star formation. A potential scenario for \gal\ is that gas is constantly accreted over time. Due to the high stability of the disc, however, \gal\ was not able to convert the accreted gas efficiently into stars as it is accreted and the massive \hi\ disc built up over time. A similar scenario is suggested for \highmass\ galaxy UGC\,12506 \citep{Hallenbeck2014} and Malin\,1 \citep{Boissier2016}.

The reason for the high specific angular momentum can only be speculated about. Possible scenarios are that either \gal\ has been formed in a dark matter halo at the high-spin tail of the halo spin distribution, \gal\ has accreted angular momentum e.\,g. through accretion from filaments that were aligned with the galaxy rotation or \gal\ simply has not lost any angular momentum over time. The semi-analytic simulation Dark-SAGE \citep{Stevens2016} suggest that the angular momentum of galaxies in the local universe is not determined by the angular momentum of the host halo. Current hydrodynamic simulations suggest that galactic discs lose angular momentum in major mergers \citep{Lagos2016} and galactic winds increase the angular momentum \citep{Genel2015}. As \gal\ is very isolated it might have never encountered any major merger, in which it could have lost angular momentum. It might have furthermore gone through an epoch of very active star formation, which in turn will have increased galactic winds and the angular momentum of the disc. 

At the current SFR and without any further gas accretion, it would take \gal\ approximately 15\,Gyr to return onto the \citet{Denes2014} scaling relation. So, unless, the \hi\ disc of \gal\ gets significantly stripped, it is likely to continue be an \hi\ excess galaxy for a very long time.

\subsection{HIX galaxies in context}

The \hix\ galaxies have been selected to have a high \hi\ mass in comparison to their \rband\ luminosity. For most \hix\ galaxies, this also implies a high \hi\ mass fraction at a given stellar mass. 

Comparing the \hi\ based star formation efficiency of the \hix\ and control sample suggests that \hix\ galaxies form stars systematically less efficiently than the control sample.

The stellar mass range, the morphology and the specific SFR (sSFR) of \hix\ galaxies is similar to these properties of the control galaxies. This is important as \cite{Brown2015} have identified the \hi\ mass fraction as a primary driver for sSFR as well as a residual relation with morphology and stellar mass. So, if the control and the \hi\ rich sample match in these properties, the difference in \hi\ content arises from other causes. Therefore, the comparison of this control to the \hix\ sample will enable us to investigate possible recent gas accretion events and answer the questions on where the excess gas has come from and why the excess gas has not yet been converted into stars. 

We will address these questions in subsequent papers, in which we will investigate the \hi\ disc and the distribution of the gas-phase oxygen abundance of the \hix\ galaxies in greater detail.  

\section{Conclusions}
\label{sec:conclude}

In this work we present the \hix\ galaxy survey targeting the most \hi\ massive galaxies for their \rband\ luminosity. We find that these galaxies are normal star forming spirals for their stellar mass. However, their \hi\ disc is more massive than in a control sample and they are inefficient at converting this gas reservoir into stars. 

The most extreme galaxy in the sample is \gal. Making use of tilted ring fits to ATCA \hi\ interferometric data combined with gas-phase metallicity information from  \wifes\ optical IFU spectra, we find that this galaxy has not recently undergone major or violent accretion events but appears to be accreting at low levels. Its \hi\ excess can, however, be attributed to a very high specific angular momentum, which prevents accreted gas from being transported to the centre of the disc and converted into stars. Comparing the current star formation rate of \gal\ to the amount of excess gas suggests that unless \gal\ loses \hi\ in other processes than star formation at the current rate, it will continue be an \hi\ excess galaxy for a very long time. 

\section*{Acknowledgments}
We would like to thank the anonymous referee for helpful comments that improved the paper. 

We would like to thank K. Gereb, D. Obreschkow, A. R. H. Stevens, D. Vohl, K. Glazebrook, M. Duree, J. Wang and Y. Bahe for insightful discussions and S. Forsayeth for outstanding computer support.

This research bases on data taken with the Australian Telescope Compact Array which is funded by the Commonwealth of Australia for operation as a National Facility managed by CSIRO. This is a large project with the project number C2705. We would like to thank the ATCA staff for excellent support during the observations.

KAL is supported by a CSIRO OCE Science Leader funded PhD top-up scholarship.

BC is the recipient of an Australian Research Council Future Fellowship (FT120100660). BC and LC acknowledge support from the Australian Research Council's Discovery Projects funding scheme (DP150101734).

The Parkes telescope is part of the Australia Telescope which is funded by the Commonwealth of Australia for operation as a National Facility managed by CSIRO. 

This research has made use of the VizieR catalogue access tool, CDS, Strasbourg, France.

This publication makes use of data products from the Wide-field Infrared Survey Explorer, which is a joint project of the University of California, Los Angeles, and the Jet Propulsion Laboratory/California Institute of Technology, funded by the National Aeronautics and Space Administration.

This publication makes use of data products from the Two Micron All Sky Survey, which is a joint project of the University of Massachusetts and the Infrared Processing and Analysis Center/California Institute of Technology, funded by the National Aeronautics and Space Administration and the National Science Foundation.

This publicatio is based on observations made with the NASA Galaxy Evolution Explorer. GALEX is operated for NASA by the California Institute of Technology under NASA contract NAS5-98034. 

This research has made use of SAOImage DS9, developed by Smithsonian Astrophysical Observatory. 

This research has further made use of python packages, which are provided by the community. Not explicitly mentioned in the main body of this work are: astropy \citep{AstropyCollaboratio2013}, NumPy\footnote{http://www.numpy.org/} \citep{vanderWalt2011}, SciPy\footnote{https://www.scipy.org/} \citep{Jones2001}, matplotlib\footnote{http://matplotlib.org/} \citep{Hunter2007} and APLpy\footnote{https://aplpy.github.io/}.

\bibliographystyle{mnras}
\bibliography{hix-1_main}

\bsp

\label{lastpage}

\end{document}